\documentclass[prb,aps,showpacs,twocolumn,floatfix]{revtex4}
\usepackage{amsmath,amsfonts,amssymb,graphics,graphicx,color,bm}
\newcommand{\bk}{{\bm k}}
\newcommand{\bS}{{\bm S}}
\begin{document}

\title{Hierarchical mean-field approach to the
       $J_1$-$J_2$ Heisenberg model on a square lattice}
\author{L. Isaev$^1$}
\author{G. Ortiz$^1$}
\author{J. Dukelsky$^2$}
\affiliation{$^1$Department of Physics, Indiana University, Bloomington IN
47405, USA \\
$^2$Instituto de Estructura de la Materia - CSIC, Serrano 123,
28006 Madrid, Spain}
\begin{abstract}
 We study the quantum phase diagram and excitation spectrum of the
 frustrated $J_1$-$J_2$ spin-1/2 Heisenberg Hamiltonian. A hierarchical
 mean-field approach, at the heart of which lies the idea of identifying
 {\it relevant} degrees of freedom, is developed. Thus, by performing
 educated, manifestly symmetry preserving mean-field approximations, we
 unveil fundamental properties of the system. We then compare various
 coverings of the square lattice with plaquettes, dimers and other
 degrees of freedom, and show that only the {\it symmetric plaquette}
 covering, which reproduces the original Bravais lattice, leads to the
 known phase diagram. The intermediate quantum paramagnetic phase is
 shown to be a (singlet) {\it plaquette crystal}, connected with the
 neighbouring N\'eel phase by a continuous phase transition. We also
 introduce fluctuations around the hierarchical mean-field solutions,
 and  demonstrate that in the paramagnetic phase the ground  and first
 excited states are separated by a finite gap, which closes in the
 N\'eel and columnar phases. Our results suggest that the quantum phase
 transition between N\'eel and paramagnetic phases can be properly
 described within the Ginzburg-Landau-Wilson paradigm.
\end{abstract}
\maketitle

\section{Introduction}

One of the primary goals of modern condensed matter physics is the
characterization of strongly correlated quantum systems. A large class
of such materials is represented by frustrated antiferromagnets, which
are believed to exhibit a variety of novel states of matter at
sufficiently strong coupling. Growing experimental evidence indicates 
that layered materials such as  ${\rm Li_2VO(Si,Ge)O_4}$ \cite{Millet},
${\rm VOMoO_4}$ \cite{Carretta} and ${\rm BaCdVO(PO_4)_2}$ \cite{Nath}
can be adequately described by an antiferromagnetic Heisenberg model
with frustrating next- and next-next-nearest neighbor interactions. As a
result, the study of low-dimensional magnets and their
frustration-driven quantum phase transitions have attracted a lot of
theoretical attention in the last decade
\cite{Sachdev_1999,Lhuillier_2004}.

A paradigmatic system, illustrating the effects of frustrating
couplings, is the spin-$1/2$ Heisenberg model on a square lattice with
competing nearest ($J_1$), and next-nearest ($J_2$) neighbor
antiferromagnetic (AF) interactions ($J_1$-$J_2$ model). Despite
numerous analytical and numerical efforts, its phase diagram, which
exhibits a two sublattice N\'eel AF, quantum paramagnetic, and a four
sublattice columnar AF states, continues to stir certain controversy
(for a review of recent achievements, see Ref.
\onlinecite{Lhuillier_2004}). While existence of the N\'eel-ordered
phase at small frustration ratio $J_2/J_1$, and of the columnar AF state
at large $J_2/J_1$ is widely established, properties of the
intermediate non-magnetic phase, which occurs around the maximum
frustration value $J_2/J_1=1/2$, are still under debate. Particularly,
the correlated nature  of the intermediate state and the kind of quantum
phase transition separating it from the N\'eel state, attract most
attention. Various methods have been recently applied to characterize
the quantum paramagnetic phase, such as Green's function Monte Carlo
\cite{Sorella_2000,Sorella_2001,Leeuwen_2000}, coupled cluster methods
\cite{Richter_2008}, series expansions \cite{Singh_1999} and
field-theoretical methods \cite{Takano_2003,Kotov_1999,Parola_2006}. As
a result, several possible candidate ground states were proposed,
namely: spin liquid \cite{Sorella_2001}, preserving translational and
rotational symmetries of the lattice, as well as various lattice
symmetry breaking phases, out of which the dimer
\cite{Kotov_1999,Sachdev_1991}, and the {\it plaquette resonating
valence bond} phases \cite{Sorella_2000} are worth mentioning.

Not surprisingly, the nature of the quantum phase transition
separating the N\'eel-ordered and quantum paramagnetic phases is also
under scrutiny. The most dramatic, and at the same time original,
scenario \cite{Sachdev_2004} is believed to violate the
Ginzburg-Landau-Wilson paradigm of phase transitions \cite{Landau_1999}
which revolves around the concept of an order parameter. Such point of
view is based on the observation that there are different spontaneously
broken symmetries in the N\'eel and quantum paramagnetic phases, which
thus cannot be connected by a group-subgroup relation. The former, of
course, breaks the $SU(2)$ invariance of the Hamiltonian and lattice
translational symmetry $T$ \cite{Manousakis_1989,cluster}, but preserves
the four-fold rotational symmetry of the square $C_4$. On the other
hand, the paramagnetic phase is known to restore the spin-rotational
symmetry and is believed to break $T$ and $C_4$, due to spontaneous
formation of dimers along the links of the lattice
\cite{Sachdev_1991,Kotov_1999}. It follows then, that these two phases
can not be joined by the usual Landau second-order critical point. This
phase transition can either be of the first-order \cite{Sirker_2006}
(the latest coupled cluster calculations \cite{Richter_2008}, however,
seem to rule out this possibility), or represent an example of a second
order critical point, which cannot be described in
terms of a bulk order parameter, but rather in terms of emergent
fractional excitations (spinons), which become deconfined right at the
critical point \cite{Sachdev_2004}.

However, evidence regarding the structure of the non-magnetic phase is
quite controversial. Indeed, the results of spin-wave calculations
\cite{Kotov_1999}, large-$N$ expansions \cite{Sachdev_1991}, and
calculations using the density matrix renormalization group combined
with Monte-Carlo simulations \cite{Leeuwen_2000} are believed to
indicate the emergence of a dimer order. On the other hand, Monte-Carlo
\cite{Sorella_2000} and coupled cluster calculations
\cite{Richter_2008}, and analytical results \cite{Takano_2003} seem to
support the presence of $C_4$ symmetry (plaquette-type ordering) in the
paramagnetic phase. In the absence of a reliable numerical or analytical
proof of existence of any particular order in the non-magnetic
region, there is no apparent reason to believe in the exotic deconfined
quantum criticality scenario. Although there apparently exists numerical
evidence \cite{Sandvik_2007}, at the moment of writing the authors are
unaware of a local Hamiltonian in space dimensions larger than one,
rigorously proven to exhibit the type of  quantum critical point
discussed in Ref. \onlinecite{Sachdev_2004}. Interestingly, it was
demonstrated in Ref. \onlinecite{Batista_2004} that a two-dimensional
(2D) lattice model can possess a {\it first order} quantum critical
point, which exhibits deconfined excitations.

All in all, the complexity of methods used to infer properties of
the paramagnetic phase and the variety of different conclusions have
created a certain degree of confusion. Our goal in the present paper is
to try to clarify some of this controversy by proposing a controlled and
manifestly symmetry preserving method, geared to computing ground state
properties of the $J_1$-$J_2$ model. Our approach is based on the
recently proposed systematic methodology to investigate the behavior of
strongly coupled systems \cite{Ortiz_2003}, whose main idea consists of
identifying {\it relevant} degrees of freedom and performing an educated
approximation, called the hierarchical mean-field (HMF), to uncover the
phase diagram and other properties of the system of interest. In a
future work these ideas will be coupled to a new, variational with
respect to the energy, renormalization group approach, which thus adapts
to the concept of relevant degrees of freedom.

\begin{figure}[!t]
 \begin{center}
  \includegraphics[width=\columnwidth]{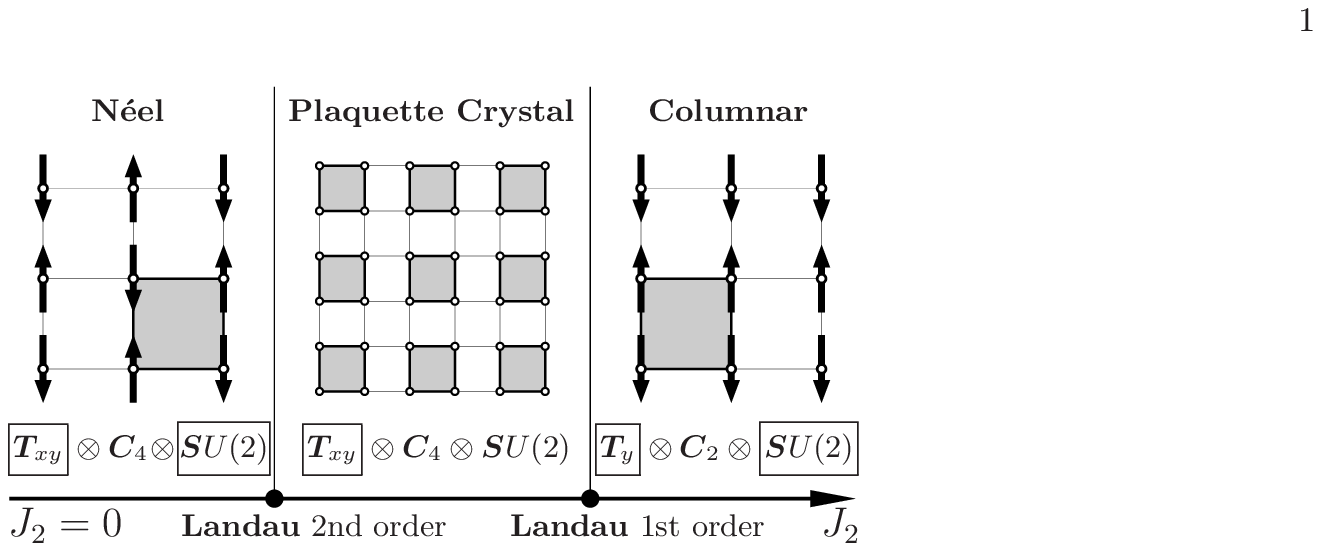}
 \end{center}
 \caption{A schematic phase diagram of the $J_1$-$J_2$ model,
	  summarizing our results.
          In each phase we show spontaneously broken (framed
	  symbols) and unbroken symmetries (usual symbols). The
	  translational invariance is broken along both directions in
	  the N\'eel and paramagnetic phases, and only along the
	  $y$--direction in the columnar phase. This fact is indicated
	  by the subscripts $xy$ and $y$ after $T$.
          Conclusions regarding the order of the phase transition,
          separating N\'eel and plaquette crystal phases, as well as symmetries
          of various phases, are based upon extrapolation of our results
          towards the thermodynamic limit.}
 \label{fig_phase_diagram}
\end{figure}

In the present work we construct HMF approximations for the $J_1$-$J_2$
model. The crux of our method is the identification of a {\it plaquette}
(spin cluster $2\times2$ or even larger $4\times4$ ({\it
superplaquette}) symmetry-preserving cluster) as the relevant elementary
degree of freedom, which captures necessary quantum correlations to
represent essential features of the phase diagram. The importance of
this degree of freedom was realized only recently in the present context
\cite{Takano_2003}, and somewhat earlier in connection with $SU(4)$
spin-orbital \cite{Mila_2000}, and Hubbard \cite{Altman_2002} models.
Besides being variational, our formalism has the attractive feature of
preserving fundamental lattice point symmetries and the $SU(2)$ symmetry
of the Hamiltonian, by utilizing the Schwinger boson-type representation
and Racah algebra technology. Remarkably, such simple mean-field
calculation already yields all known results, concerning the phase
diagram of the $J_1$-$J_2$ model, with a good accuracy, namely:
existence of a N\'eel-ordered phase with  antiferromagnetic wavevector
$(\pi,\pi)$ and spin-wave type excitations for $J_2/J_1\lesssim 0.42$, a
non-magnetic intermediate gapped phase, separated by a second order
quantum phase transition, and  a first order transition point, which is
characterized by the discontinuous disappearance of the energy gap and
connects the paramagnetic state with the columnar antiferromagnetic
phase at $(\pi,0)$ and $(0,\pi)$ for $J_2/J_1\gtrsim0.66$.

We emphasize that our investigation primarily focuses on the symmetry
analysis of the various phases. Out of many possible coarse graining
scenarios, such as covering of the 2D lattice with plaquettes, dimers
and crosses, only the $C_4$-{\it symmetry preserving} plaquette (or
superplaquette) covering (which reproduces the original Bravais lattice)
displays the correct phase diagram. In particular, the intermediate
paramagnetic phase is shown to be a {\it plaquette crystal}, which
preserves spin and lattice rotational symmetries. For all other scenarios,
including dimerized (bond-ordered) phases, we were unable to reproduce all
known quantum phase transition points of the model.

We notice that the HMF coarse graining procedure leads to an explicit breaking
of a particular translational symmetry. As a result, one can not draw rigorous
conclusions on the order of the phase transitions, based solely on a fixed
coarse graining. Nevertheless, it is still possible to make some predictions,
using a finite-size scaling of the relevant degree of freedom towards the
thermodynamic limit, where the effects associated with coarse graining should
disappear.

Next two sections are devoted to the formulation of the HMF approach. Then, we
present results of our calculations and close the paper with a discussion. Our
main conclusions are summarized in Fig. \ref{fig_phase_diagram}, which
emphasizes symmetry relations between different phases of the model.

\section{The plaquette degree of freedom}

\begin{figure}[!t]
 \begin{center}
  \includegraphics[width=0.65\columnwidth]{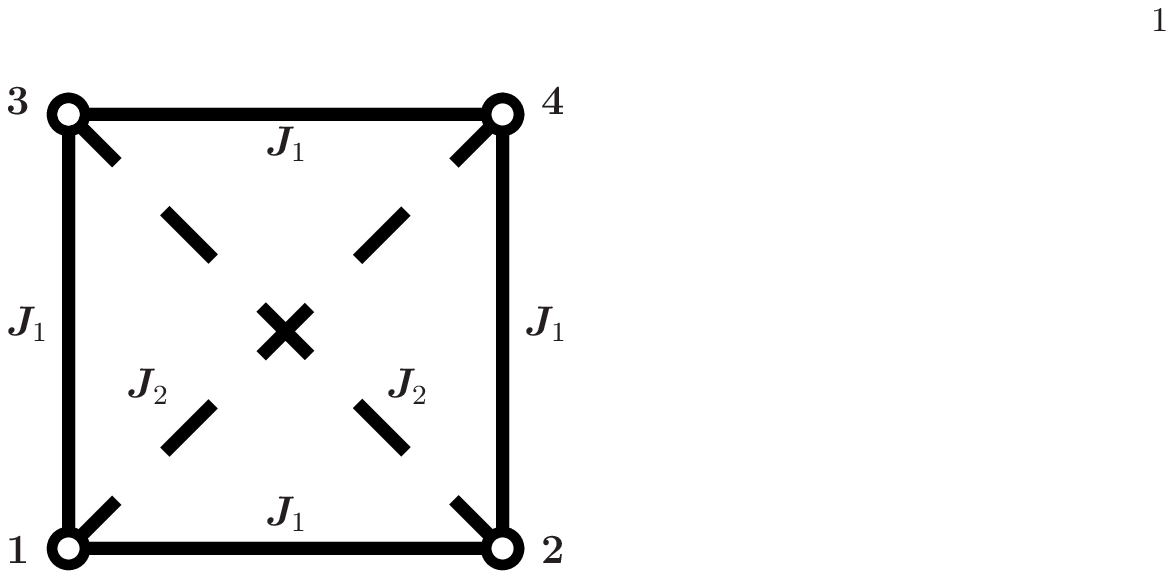}
 \end{center}
  \caption{A single $2\times2$ plaquette has each vertex occupied by a
	   $S=1/2$ spin. The diagonal spins interact through a
	   Heisenberg term of strength $J_2$, while nearest neighbor
	   spins interact with strength $J_1$.}
  \label{fig_plaquette}
\end{figure}

We consider the spin-$1/2$ antiferromagnetic Heisenberg model with
frustrated next-nearest neighbor interactions $J_2$, defined on a 2D
bipartite lattice with $N$ sites:
\begin{equation}
 H=J_1\sum_{\langle i,j\rangle}{\bm S}_i\cdot{\bm
 S}_j+J_2\sum_{\langle\langle i,j\rangle\rangle}{\bm S}_i\cdot{\bm S}_j.
 \label{hamiltonian}
\end{equation}
As mentioned already in the Introduction, we choose the plaquette, Fig.
\ref{fig_plaquette}, as our {\it elementary} degree of freedom. Then,
assuming that $N$ is chosen appropriately, the entire lattice can be
covered with such plaquettes in a sub-exponentially \cite{Mila_2000}
($\sim2^{\sqrt{N}}$) large number of ways.

\begin{figure}[!t]
 \begin{center}
  \includegraphics[width=0.8\columnwidth]{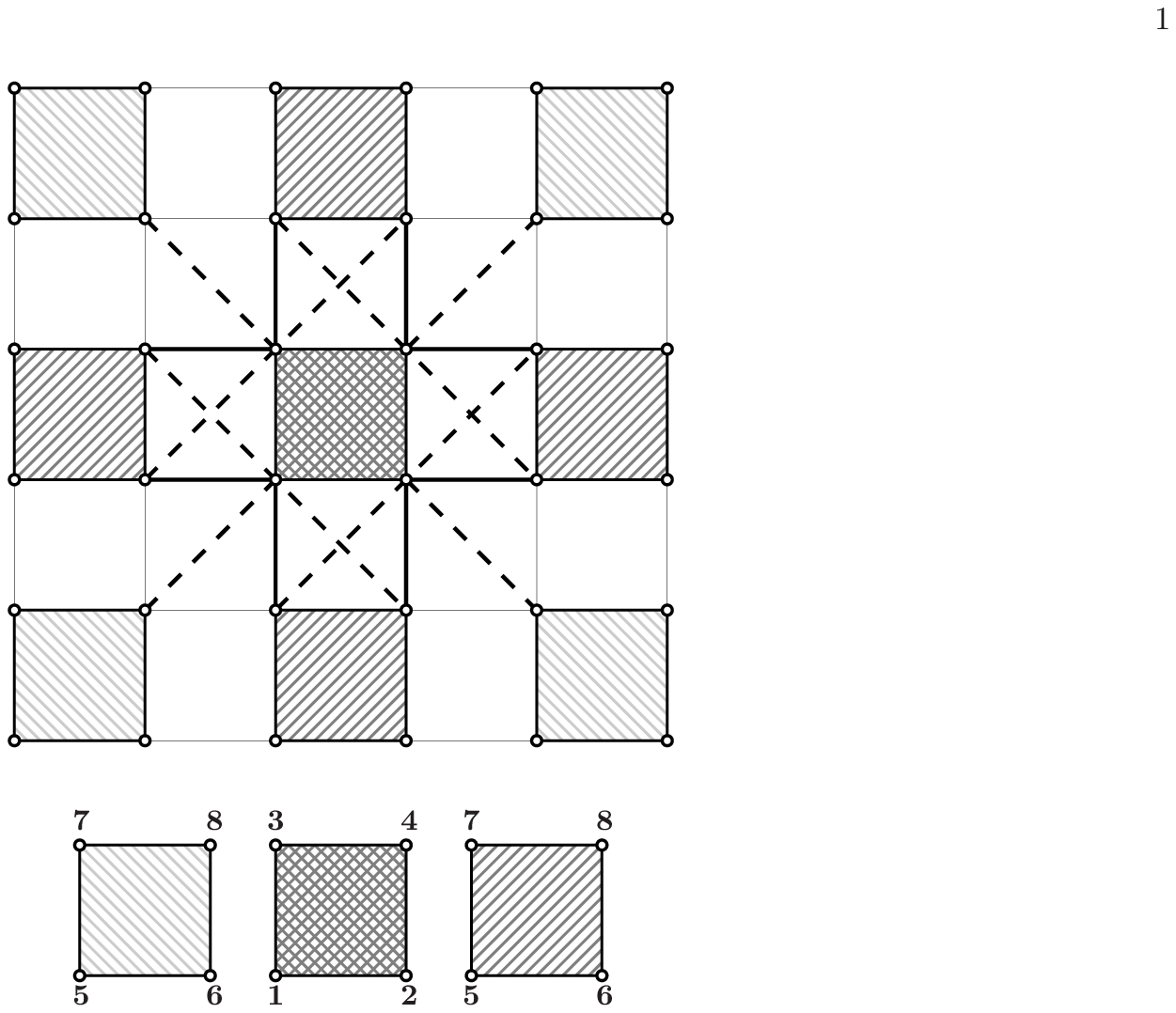}
 \end{center}
 \caption{Symmetric covering of the 2D lattice with $2\times2$
	  plaquettes. Each plaquette is connected to 4 nearest and 4
	  next-nearest neighbors.}
 \label{symm_plaquette_covering}
\end{figure}

Aiming at illustrating the main idea of the method, in this section we
consider in detail only the symmetric covering of the lattice with
$2\times2$ plaquettes, which preserves the $C_4$ lattice symmetry, see
Fig. \ref{symm_plaquette_covering}, although later the displaced
covering (Fig. \ref{displ_plaquette_covering}), which breaks $C_4$ down
to $C_2$ (two-fold symmetry axis), and the case of larger plaquettes
(superplaquettes, Fig. \ref{fig_sp}) will be analyzed as well.

It is convenient to take as a basis the states
\begin{equation}
 |a\rangle=|l_1l_2LM\rangle,\label{basis}
\end{equation}
where ${\bm l}_1={\bm S}_1+{\bm S}_4$ and ${\bm l}_2={\bm S}_2+{\bm
S}_3$ are total spins of the plaquette diagonals, while ${\bm L}={\bm
l}_1+{\bm l}_2$ is the total spin of the entire plaquette and $M$ is its
$z$-component. In this basis the Hamiltonian of a single plaquette,
\begin{equation}
 H_\Box=J_1\bigl({\bm S}_1+{\bm S}_4\bigr)\bigl({\bm S}_2+{\bm S}_3\bigr)+
 J_2\bigl({\bm S}_1\cdot{\bm S}_4+{\bm S}_2\cdot{\bm S}_3\bigr)
 \label{plaquette_hamiltonian}
\end{equation}
is diagonal with eigenvalues
\begin{align}
 \epsilon_{l_1l_2L}=&\frac{J_1}{2}\bigl[L\bigl(L+1\bigr)-l_1\bigl(l_1+1\bigr)-
 l_2\bigl(l_2+1\bigr)\bigr]+ \nonumber \\
 +&\frac{J_2}{2}\bigl[l_1\bigl(l_1+1\bigr)+l_2\bigl(l_2+1\bigr)-3\bigl].
 \label{plaquette_se}
\end{align}
We note that the basis of Eq. (\ref{basis}) is a natural one and allows
us to explicitly label states with corresponding representations of
$SU(2)$.

The next step is to establish how a plaquette couples to the rest of the
system. In Fig. \ref{symm_plaquette_covering} we show the symmetric
plaquette covering of the 2D lattice. In the figure the vertices of
every non-central plaquette are similarly labeled by the numbers
$5,6,7,8$, and total spins of diagonals are ${\bm l}_3={\bm S}_5+{\bm
S}_8$ and ${\bm l}_4={\bm S}_6+{\bm S}_7$. In the uncoupled basis matrix
elements of the inter-plaquette interaction are:
\begin{align}
 \bigl(H_{\rm int}^\sigma&\bigr)^{a_1^\prime a_2^\prime}_{a_1a_2}=
 \sum_{LM}\bigl\langle\lambda_1^\prime\lambda_2^\prime,LM|H_{\rm int}^\sigma|
 \lambda_1\lambda_2,LM\bigr\rangle\times \label{interaction} \\
 &\times\bigl\langle L_1^\prime M_1^\prime L_2^\prime M_2^\prime|
 L_1^\prime L_2^\prime LM\bigr\rangle\bigl\langle L_1M_1L_2M_2|L_1L_2LM\bigr
 \rangle, \nonumber
\end{align}
where $\sigma=1$ ($\sigma=2$) corresponds to the nearest (next-nearest)
neighbor interaction, $L_1$,$L_2$ ($L_1^\prime$,$L_2^\prime$)
represent initial (final) angular momenta of the two plaquettes and
${\bm L}={\bm L}_1+{\bm L}_2$ is their total angular momentum. In this
equation we have introduced the notations $\lambda_1=\{l_1l_2L_1\}$,
$\lambda_2=\{l_3l_4L_2\}$ and $a_i=\{\lambda_iM_i\}$, and
similarly for the primed indices. Because
each plaquette has 4 nearest neighbors and 4 next-nearest neighbors (see
Fig. \ref{symm_plaquette_covering}), the symmetrized next-nearest
neighbor interaction may be written as:
\begin{align}
 \bigl\langle&\lambda_1^\prime\lambda_2^\prime,LM|H_{\rm int}^2|\lambda_1
 \lambda_2,LM\bigr\rangle=J_2\rho_{L_1L_2}^{L_1^\prime L_2^\prime}(L)\times
 \label{c_4_nnn_int} \\
 \times&\biggl(\!S_3^{\lambda_1^\prime\lambda_1}S_6^{\lambda_2^\prime\lambda_2}
 \!\!+S_1^{\lambda_1^\prime\lambda_1}S_8^{\lambda_2^\prime\lambda_2}\!\!+
 S_2^{\lambda_1^\prime\lambda_1}S_7^{\lambda_2^\prime\lambda_2}\!\!+
 S_4^{\lambda_1^\prime\lambda_1}S_5^{\lambda_2^\prime\lambda_2}\!\biggr).
 \nonumber
\end{align}
while the symmetrized nearest neighbor plaquette interaction has the form:
\begin{align}
 \bigl\langle\lambda_1^\prime&\lambda_2^\prime,LM|H_{\rm int}^1|\lambda_1
 \lambda_2,LM\bigr\rangle=J_1\rho_{L_1L_2}^{L_1^\prime L_2^\prime}(L)\times
 \label{c_4_nn_int} \\
 &\times\bigl[\bigl(S_1^{\lambda_1^\prime\lambda_1}+S_4^{\lambda_1^
 \prime\lambda_1}\bigr)\bigl(S_6^{\lambda_2^\prime\lambda_2}+S_7^{\lambda_2^
 \prime\lambda_2}\bigr)+ \nonumber \\
 &\qquad+\bigl(S_2^{\lambda_1^\prime\lambda_1}+S_3^{\lambda_1^
 \prime\lambda_1}\bigr)\bigl(S_5^{\lambda_2^\prime\lambda_2}+S_8^{\lambda_2^
 \prime\lambda_2}\bigr)\bigr]+ \nonumber \\
 &\qquad+2\bigl\langle\lambda_1^\prime\lambda_2^\prime,LM|H_{\rm int}^2|
 \lambda_1\lambda_2,LM\bigr\rangle, \nonumber
\end{align}
In Eqs. (\ref{c_4_nnn_int}) and (\ref{c_4_nn_int}) the symbols
$S_n^{\lambda^\prime\lambda}=\langle\lambda^\prime\|S_n\|\lambda\rangle$
denote reduced matrix elements of the $n$-th spin operator, and:
\begin{displaymath}
 \rho_{L_1L_2}^{L_1^\prime L_2^\prime}(L)=\frac{1}{4}(-1)^{L+L_2^\prime+L_1}
 \left\{
 \begin{array}{ccc}
  L_1^\prime & L_2^\prime & L \\
  L_2 & L_1 & 1
 \end{array}
 \right\} ,
\end{displaymath}
where $\{\cdots\}$ are Wigner $6j$ symbols (or Racah coefficients)
\cite{Edmonds_1957}.

Let us now identify the plaquette degree of freedom with a Schwinger boson
which creates a specific state of the plaquette. Then, the Hamiltonian of Eq.
(\ref{hamiltonian}) in the plaquette basis can be expressed as:
\begin{align}
 H=&\sum_{i,a}\epsilon_{a}\gamma_{ia}^\dag\gamma_{ia}+\sum_{\langle ij\rangle}
 \bigl(H_{\rm int}^1\bigr)^{a_1^\prime a_2^\prime}_{a_1a_2}\gamma_{ia_1^
 \prime}^\dag\gamma_{ja_2^\prime}^\dag\gamma_{ia_1}\gamma_{ja_2}+ \nonumber \\
 &+\sum_{\langle\langle ij\rangle\rangle}\bigl(H_{\rm int}^2\bigr)^{a_1^\prime
 a_2^\prime}_{a_1a_2}\gamma_{ia_1^\prime}^\dag\gamma_{ja_2^\prime}^\dag\gamma_
 {ia_1}\gamma_{ja_2},
 \label{ham_second_quant}
\end{align}
where the operator $\gamma_{ia}^\dag$ creates a boson on site $i$ of the
{\it plaquette} lattice (which contains $N_\Box=N/4$ sites) in the
state, denoted by an index $a$, running through the entire
single-plaquette Hilbert space (of dimension $2^4=16$) and the summation
is performed over doubly repeated dummy indices. The unphysical states
are eliminated by enforcing the local constraint
$\sum_a\gamma_{ia}^\dag\gamma_{ia}=1$. In what follows, we impose
periodic boundary conditions on the plaquette lattice.

The bosonic operators $\gamma_{ia}$ define the hierarchical language
\cite{Ortiz_2003} for our problem. It will be used in the next section,
where we develop an approximation scheme for diagonalizing the
Hamiltonian of Eq. (\ref{ham_second_quant}).

\section{Hierarchical mean-field approximation}

As it follows from Eq. (\ref{plaquette_se}), the lowest single-plaquette
state has the energy $\epsilon_{1100}/4=-J_1/2+J_2/8$ per spin, which,
when $J_2=0$, gives only the energy of a classical 2D antiferromagnet.
Thus, it is necessary to take into account the interaction term in Eq.
(\ref{ham_second_quant}).

The HMF approximation is a mean-field approach, performed on the
relevant degrees of freedom. In the present section we discuss only the
simplest one -- a Hatree-Fock like (HF) approximation. A possible way to
include fluctuation corrections is presented in Appendix B. The HF
approximation introduces the mixing of single-plaquette states which
minimizes the  total energy of the system and is based on a canonical
transformation among the bosons, which we will restrict to be uniform
(plaquette independent):
\begin{equation}
 \gamma_{ia}=R_a^n\Gamma_{in}.
\end{equation}
The real matrix $R$ satisfies canonical orthogonality and
completeness relations:
\begin{displaymath}
 R_a^nR_a^{n^\prime}=\delta_{nn^\prime};\,\,\,R_a^nR_{a^\prime}^n=
 \delta_{aa^\prime}
\end{displaymath}
A {\it translationally invariant} variational ansatz for the ground
state (vacuum) is a boson condensate in the lowest HF single-particle
energy state ($n=0$):
\begin{equation}
 |{\rm HF}\rangle=\prod_i\Gamma_{i0}^\dag|0\rangle,\label{hf_wavefunction}
\end{equation}
and since it has one boson per plaquette, there is no need to impose the
Schwinger boson constraint in the calculation.

Minimizing the total energy with respect to $R$, we arrive at the
self-consistent equation:
\begin{equation}
 \biggl\{\epsilon_a\delta_{aa^\prime}+\sum_{\sigma}z_\sigma\bigl(H_{\rm int}^
 \sigma\bigr)^{aa_1}_{a^\prime a_2}R_{a_1}^0R_{a_2}^0\biggr\}R_{a^\prime}^n=
 \varepsilon_nR_a^n,\label{HMF_equation}
\end{equation}
where $z_1=z_2=4$ are the nearest- and next-nearest coordination numbers. The
ground state energy (GSE) per spin is then given by the expression:
\begin{align}
 \frac{E_0}{N}=\frac{\langle{\rm HF}|H|{\rm HF}\rangle}{N}=\frac{1}{8}\biggl(
 \varepsilon_0+\sum_{a}\epsilon_a\bigl(R_a^0\bigr)^2\biggr) , 
 \label{HMF_energy}
\end{align}
with $\varepsilon_0$ being the lowest eigenvalue of Eq. (\ref{HMF_equation}).

Another fundamental quantity to compute is the polarization of spins within a
plaquette:
\begin{displaymath}
 \langle{\rm HF}|S_{in}^z|{\rm HF}\rangle=\left(S_n^z\right)_{a^\prime a}
 R^0_{a^\prime}R^0_a,
\end{displaymath}
where $n=1,\dots,4$ is the spin index, and the matrix elements (determined
from the Wigner-Eckart theorem) are:
\begin{align}
 &\left(S_n^z\right)_{a^\prime a}=\langle l_1^\prime l_2^\prime L^\prime
 M^\prime|S_n^z|l_1l_2LM\rangle= \label{spin_matr_el} \\
 =&(-1)^{L+L^\prime+1}\delta_{MM^\prime}\frac{\langle10LM|1LL^\prime M\rangle}
 {\sqrt{2L^\prime+1}}\langle l_1^\prime l_2^\prime L^\prime||S_n||l_1l_2L
 \rangle.  \nonumber
\end{align}
This enables us to define the staggered and collinear (along $x$ and $y$ axes)
magnetizations:
\begin{align}
 M_{\rm stag}&=(1/4)\langle {\rm HF}|S_1^z+S_4^z-S_2^z-S_3^z|{\rm HF}\rangle;
 \label{magnetizations} \\
 M_{\rm col}(x,y)&=(1/4)\langle {\rm HF}|S_1^z-S_4^z+S_{2,3}^z-S_{3,2}^z|
 {\rm HF}\rangle. \nonumber
\end{align}

Notice the extreme simplicity of the HMF approximation. The reason why
it is able to realize meaningful results is that the plaquette degree of
freedom seems to contain the main correlations defining the physics
behind the Hamiltonian of Eqs. (\ref{hamiltonian}), and
(\ref{ham_second_quant}). To avoid confusion, we emphasize that the HF
approximation and the fluctuation theory of Appendix B are physically
(and obviously mathematically) different from the spin-wave or canonical
Schwinger-Wigner boson mean-field approach to spin systems
\cite{Auerbach_1994}. In particular, we make no assumption about the
underlying ground state, thus allowing for an interplay of various
quantum phases. Moreover, it will be demonstrated, that the collective
excitation spectra in each phase consistently reflect spontaneously
broken symmetries, unlike the usual Schwinger boson case
\cite{Auerbach_1994}, in which one obtains gapped excitations.

\section{Ground state properties and excitation spectrum of the model}

\begin{figure}[!t]
 \begin{center}
  \includegraphics[scale=0.3]{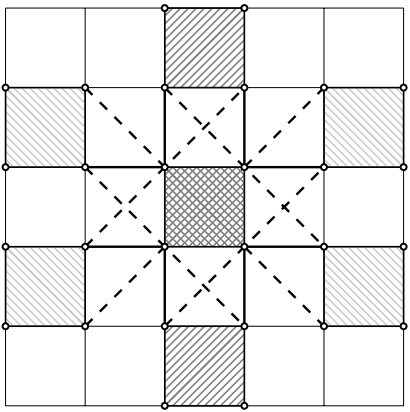}
 \end{center}
 \caption{The displaced plaquette covering. Notice that the $C_4$
	  symmetry is broken down to $C_2$.}
 \label{displ_plaquette_covering}
\end{figure}
\begin{figure}[!t]
 \begin{center}
  \includegraphics[scale=0.3]{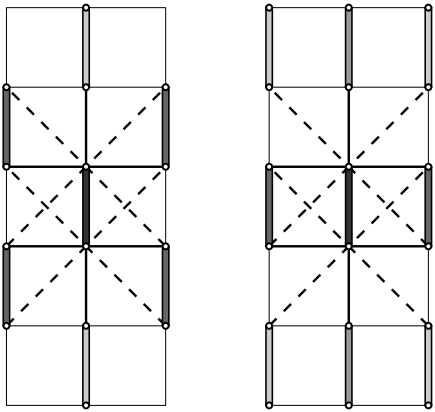}
 \end{center}
 \caption{Connectivity of the dimer lattice for symmetric (right panel)
	  and displaced dimer coverings (left panel). The rotational
	  $C_4$ symmetry is lowered to $C_2$ in both cases.}
 \label{dimer_covering}
\end{figure}

\begin{figure}[!t]
 \begin{center}
  \includegraphics[scale=0.3]{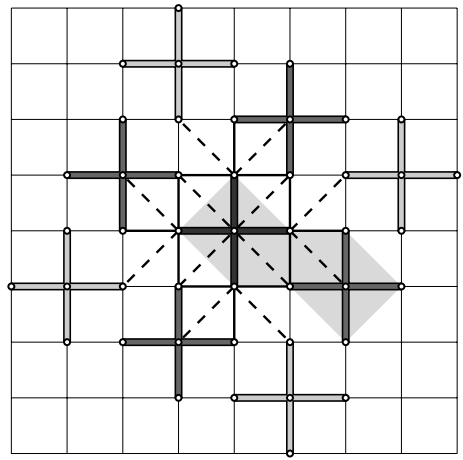}
 \end{center}
 \caption{Covering of the lattice with crosses -- arrays of five spins.
	  Since one cross cannot form a singlet, it is necessary to
	  double the unit cell, as indicated by the gray shading. This
	  choice of a degree of freedom clearly preserves the $C_4$
	  symmetry,  but the resulting lattice breaks it.}
 \label{cross_covering}
\end{figure}

Our choice of the plaquette as an elementary degree of freedom remains
unjustified at this point. In order to show its relevance we applied the
analysis of two previous sections to several other coarse grainings
(besides the symmetric plaquette covering, case (a), shown in Fig.
\ref{symm_plaquette_covering}): (b) {\it superplaquette} (spin cluster
$4\times4$) degree of freedom, covering the lattice in such a way that
$C_4$ is preserved (see Appendix A for details); (c) displaced plaquette
covering of the lattice, Fig. \ref{displ_plaquette_covering}; (d)
symmetric and displaced {\it dimer} coverings, shown respectively in
right and left panels of Fig. \ref{dimer_covering}; (e) {\it cross}
degree of freedom, Fig. \ref{cross_covering}. One should observe that 
symmetries of the original Bravais lattice are preserved only in cases
(a) and (b). In cases (c) and (d) the lattice rotational symmetry $C_4$
is lowered to $C_2$. Case (e) is special in the sense that an isolated
degree of freedom does not possess a singlet ground state. The
information about a particular configuration is encoded in matrix
elements of $H_{\rm int}^\sigma$, whose calculation is elementary. Other
equations, presented in Secs. II and III, retain their form.

For each of the above cases we iteratively solve Eq. (\ref{HMF_equation}) and
compute the GSE (\ref{HMF_energy}), and  staggered and collinear
magnetizations (\ref{magnetizations}). The main message, which we would
like to convey in this section is that only the plaquette degree of
freedom (of any size) is relevant for constructing the phase diagram of
the Hamiltonian of Eq. (\ref{hamiltonian}).

\subsection{Symmetry preserving plaquette configurations}

\begin{figure}[!t]
 \begin{center}
  \includegraphics[width=\columnwidth]{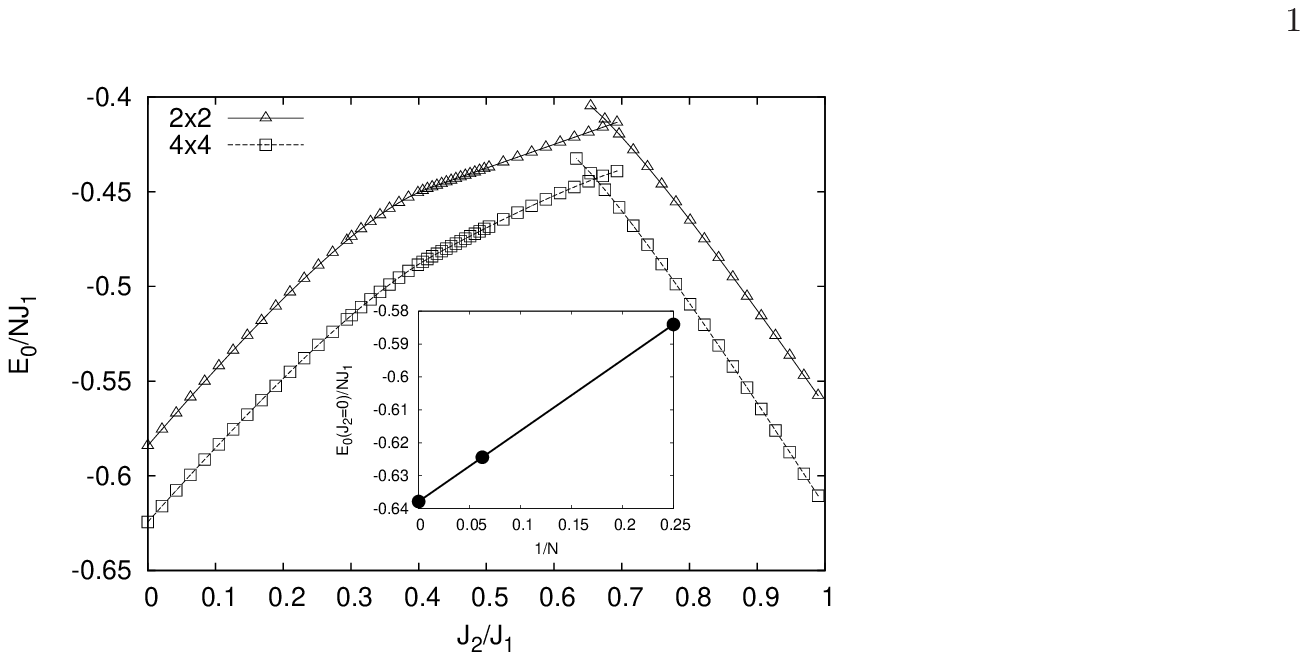}
 \end{center}
 \caption{Ground state energy per spin computed at the HF level for the
	  $2\times2$ and $4\times4$ plaquette elementary degrees of
	  freedom. The inset shows finite-size scaling in the AF phase
          at $J_2=0$.}
 \label{fig_gse_2x2_4x4}
\end{figure}

\begin{figure}[!t]
 \begin{center}
  \includegraphics[width=\columnwidth]{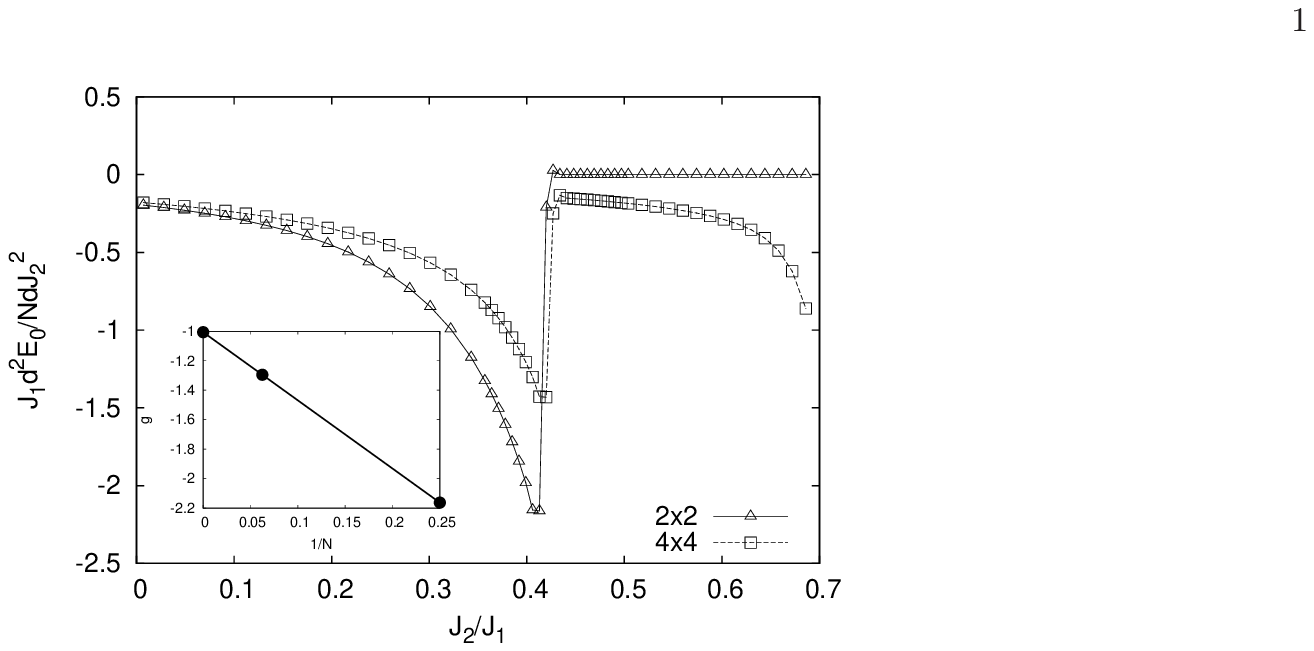}
 \end{center}
 \caption{Second-order derivative $d^2E_0/dJ_2^2$ for the $2\times2$ and
	  $4\times4$ plaquette degrees of freedom. The discontinuity at
	  $J_2/J_1\approx0.42$ is indicative of a second-order quantum
	  phase transition. In the inset we present finite-size scaling for
          the jump
          $g\equiv\left(J_1d^2E_0/NdJ_2^2\right)_{J_2^{c1}-0}^{J_2^{c1}+0}$.}
 \label{fig_gse_d2_2x2_4x4}
\end{figure}

\begin{figure}[!t]
 \begin{center}
  \includegraphics[width=\columnwidth]{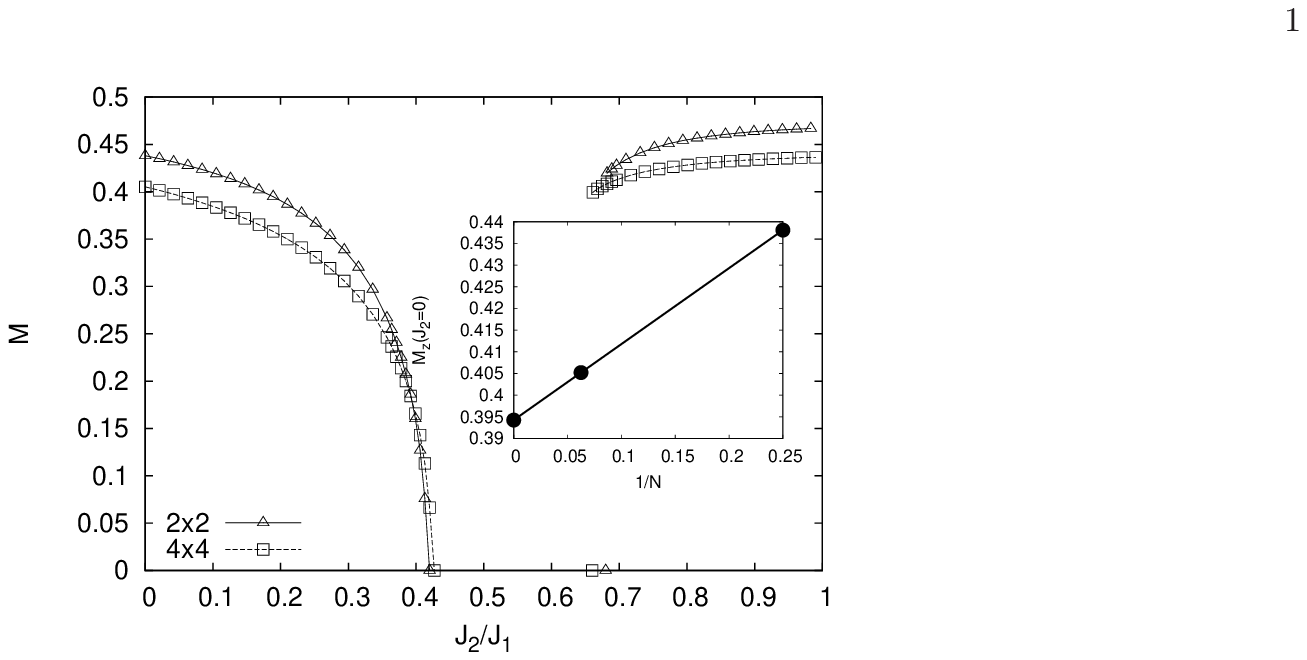}
 \end{center}
 \caption{Staggered magnetization, $M_{\rm stag}$, for $J_2\le J_2^{c1}$
	  and collinear magnetization along the $x$-direction, $M_{\rm
	  col}(x)$, for $J_2\ge J_2^{c2}$ (for the $2\times2$ and
	  $4\times4$ plaquette degrees of freedom), computed at the HF
	  level. Notice the continuous phase transition at
	  $J_2/J_1\approx0.42$ and a first order transition at
	  $J_2/J_1\approx0.68$ ($2\times2$) and $J_2/J_1\approx0.66$
	  ($4\times4$). The inset shows finite-size scaling of $M_z$ at
          $J_2=0$.}
 \label{fig_mag_2x2_4x4}
\end{figure}

Let us focus first on cases (a) and (b), i.e. symmetry-preserving
coverings of the lattice with plaquette, Fig. \ref{symm_plaquette_covering},
and superplaquette degrees of freedom. The resulting GSE as a function
of $J_2/J_1$ is shown in Fig. \ref{fig_gse_2x2_4x4}. One immediately
observes a level-crossing at $J_2^{c2}\approx0.67J_1$, indicating the
first-order transition and a second-order quantum critical point at
$J_2^{c1}\approx0.42J_1$, which is supported by a jump of the second
order derivative $d^2E_0/dJ_2^2$, Fig. \ref{fig_gse_d2_2x2_4x4}. Both
N\'eel and columnar phases are characterized by spontaneously broken
$SU(2)$ symmetry. The former exhibits a nonvanishing staggered
magnetization, $M_{\rm stag}$, while the latter has nonzero collinear
magnetization along the $x$-direction, $M_{\rm col}(x)$. Both order
parameters become zero in the paramagnetic phase, suggesting that $SU(2)$
is restored. These results are summarized in Fig. \ref{fig_mag_2x2_4x4},
from which it also follows that the phase transition at $J_2^{c1}$ is
continuous, while $J_2^{c2}$ corresponds to a first-order transition
point. We remind, in this connection, that our approach does not
explicitly break the spin rotational symmetry, thus allowing for the
treatment of competing ground states.

As expected, considering a larger elementary degree of freedom --
superplaquette -- leads to a significant improvement of the GSE and
reduction of the magnetization, due to larger quantum fluctuations. The
finite-size scaling (insets in Figs. \ref{fig_gse_2x2_4x4} and
\ref{fig_mag_2x2_4x4}), using these two sizes ($2\times2$ and
$4\times4$), indicates that $E_0(J_2=0)/N\to-0.64J_1$ and
$M_z(J_2=0)\to0.39$ in the thermodynamic limit, a satisfying result for
a HF approximation, which completely ignores fluctuations (these numbers
should be compared with well-known results of Monte-Carlo simulations
\cite{Ceperly_1989}: $E_0/N\approx-0.67J_1$ and $M_z\approx0.31$).

Next, we discuss in more detail symmetry properties of the various
phases in Figs. \ref{fig_gse_2x2_4x4}, \ref{fig_mag_2x2_4x4}. At all
values of $J_2\le J_2^{c2}$, the lattice translational symmetry $T$ is
broken \cite{cluster}, but the rotational $C_4$ symmetry is preserved.
For $J_2\le J_2^{c1}$ this corresponds to a N\'eel-type long-range order
with spontaneously broken $SU(2)$. At large values $J_2\ge J_2^{c2}$, we
observe the columnar ordering, which spontaneously breaks $C_4$ down to
$C_2$, and $SU(2)$, but partially (i.e., along one direction) restores
the lattice translational symmetry. We present a more detailed
discussion of the spatial symmetries later in this section. In the
intermediate region $J_2\in(J_2^{c1},J_2^{c2})$ the spin $SU(2)$
rotational symmetry is restored. In this paramagnetic phase the ground
state wavefunction is a tensor product of individual plaquette ground
states (with quantum numbers $l_1=l_2=1$, $L=M=0$):
\begin{align}
 |1100\rangle=\frac{1}{2\sqrt{3}}\biggl[&2\bigl(|\uparrow_1\uparrow_4
 \downarrow_2\downarrow_3\rangle+|\downarrow_1\downarrow_4\uparrow_2\uparrow_3
 \rangle\bigr)-\nonumber \\
 &-\bigl(|\uparrow_1\downarrow_4\uparrow_2\downarrow_3\rangle+|\uparrow_1
 \downarrow_4\downarrow_2\uparrow_3\rangle+ \label{plaquette_GS} \\
 &\,\,\,\,\,\,\,\,+|\downarrow_1\uparrow_4\uparrow_2\downarrow_3\rangle+
 |\downarrow_1\uparrow_4\downarrow_2\uparrow_3\rangle\bigr)\biggr]. \nonumber
\end{align}
This ground state necessarily breaks the lattice translational symmetry,
but preserves $C_4$. In fact, the paramagnetic region on the phase
diagram of Figs. \ref{fig_gse_2x2_4x4} and \ref{fig_mag_2x2_4x4} is a
trivial plaquette crystal: a set of {\it non-interacting} plaquettes,
because the expectation value of the plaquette interaction (see Eq.
(\ref{ham_second_quant})) in the singlet state $\prod_i
\gamma^\dag_{i,1100}|0\rangle$ vanishes. An analogous situation is
realized when the superplaquette is chosen as an elementary degree of
freedom: the paramagnetic phase is a crystal of superplaquettes. It is
interesting to note that in Ref. \onlinecite{Sorella_2000} a ``plaquette
resonating-valence-bond state'', exactly equal to (\ref{plaquette_GS}), has
been proposed.
However, later \cite{Sorella_2001} the intermediate phase was argued to
be a spin liquid, i.e., a state that preserves the lattice translational
symmetry.

In order to learn about spatial symmetries in various phases, we compare
magnitudes of the several lattice symmetry-breaking observables proposed
in the literature. We consider the following three, introduced in Ref.
\onlinecite{Richter_2008} (in the notation of that paper):
\begin{align}
 F_1=&\frac{1}{N}\sum_{x,y}(-1)^x{\bm S}_{x,y}{\bm S}_{x+1,y}; \nonumber \\
 F_2=&\frac{1}{N}\sum_{x,y}{\bm S}_{x,y}\bigl({\bm S}_{x+1,y}-{\bm S}_{x,y+1}
 \bigr); \nonumber \\
 F_4=&\frac{1}{N}\sum_{x,y}{\bm S}_{x,y}\bigl[(-1)^x{\bm S}_{x+1,y}+(-1)^y{\bm
 S}_{x,y+1}\bigr], \nonumber
\end{align}
where indices $x,y$ specify a spin in the 2D lattice. The operator $F_4$
probes the plaquette ordering, which preserves the lattice rotational
symmetry, while $F_1$ and $F_2$ correspond to the columnar ordering. We
note, however, that $F_1$ is already non-zero for an isolated plaquette (or
superplaquette). These functions can be combined in the complex ``order
parameter'', introduced in Ref. \onlinecite{Sachdev_2004}. Here we show
details of the calculation of functions $F_{1,2,4}$ for the plaquette degree of
freedom, case (a), and only present the result for $F_4$ for the superplaquette
case (b). In the plaquette representation the above operators are written as:
\begin{align}
 F_1=&\frac{1}{N}\sum_{i,j}\bigl[\bigl({\bm S}_{1;i,j}{\bm S}_{2;i,j}+
 {\bm S}_{3;i,j}{\bm S}_{4;i,j}\bigr)- \nonumber \\
 &\,\quad\qquad-\bigl({\bm S}_{2;i,j}{\bm S}_{5;i+1,j}+{\bm S}_{4;i,j}
 {\bm S}_{7;i+1,j}\bigr)\bigr]; \nonumber \\
 F_{2,4}=&\frac{1}{N}\sum_{i,j}\bigl[\bigl({\bm S}_{1;i,j}\mp{\bm S}_{4;i,j}
 \bigr)\bigl({\bm S}_{2;i,j}\mp{\bm S}_{3;i,j}\bigr)\pm
 \label{order_parameters} \\
 &\,\quad\qquad\pm\bigl({\bm S}_{2;i,j}{\bm S}_{5;i+1,j}+{\bm S}_{4;i,j}
 {\bm S}_{7;i+1,j}\mp \nonumber \\
 &\,\,\qquad\qquad\mp{\bm S}_{3;i,j}{\bm S}_{5;i,j+1}\mp{\bm S}_{4;i,j}
 {\bm S}_{6,i,j+1}\bigr)\bigr]. \nonumber
\end{align}
In this equation the indices $i,j$ are coordinates of a plaquette in the
lattice.

\begin{figure}[!t]
 \begin{center}
  \includegraphics[width=\columnwidth]{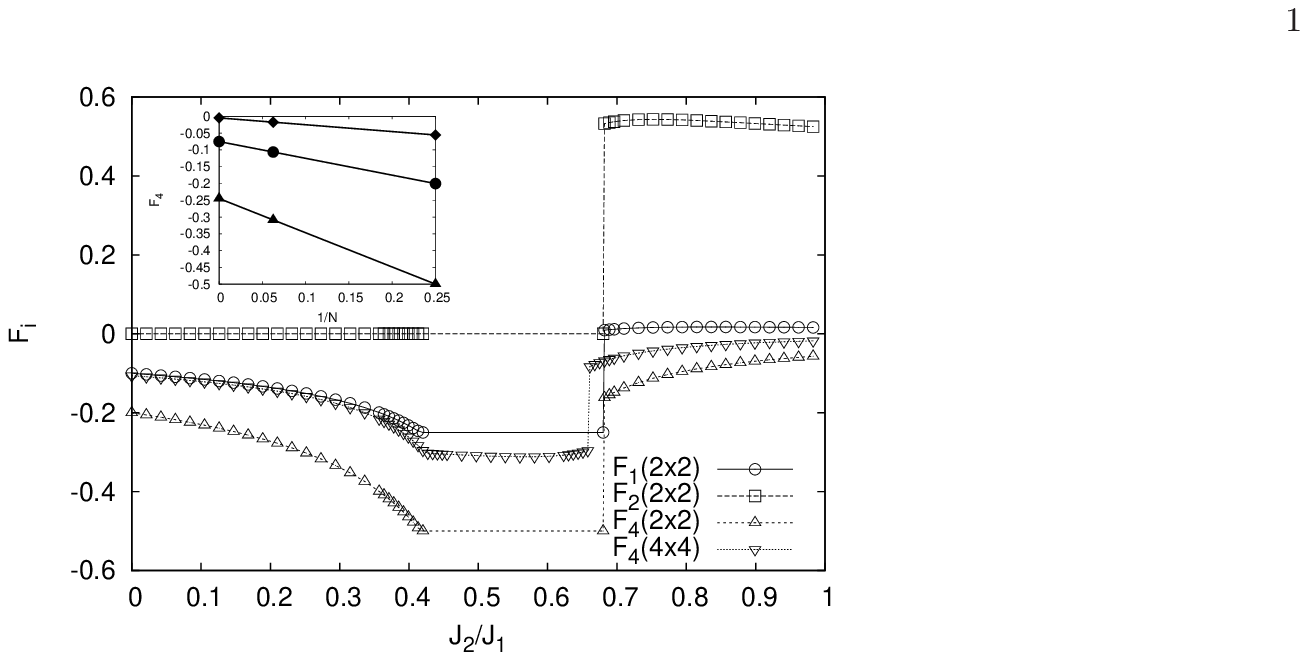}
 \end{center}
 \caption{HF ground state expectation values of the symmetry-breaking
	  perturbations, given by Eq. (\ref{order_parameters}), plotted
	  as functions of $J_2/J_1$ for  the plaquette and superplaquette
          degrees of freedom. Due to the unbroken $C_4$ symmetry in the N\'eel
	  and paramagnetic phases, values of $F_4$ are twice larger than
	  the  corresponding values of $F_1$, except in the columnar
	  phase, where $C_4$ is spontaneously broken down to $C_2$. The inset
          shows finite-size scaling for $F_4$ for three values of $J_2/J_1$:
          $0$ (circles), $0.504$ (triangles) and $0.997$ (rhombs).}
 \label{fig_order_parameters}
\end{figure}

\begin{figure}[!t]
 \begin{center}
  \includegraphics[width=\columnwidth]{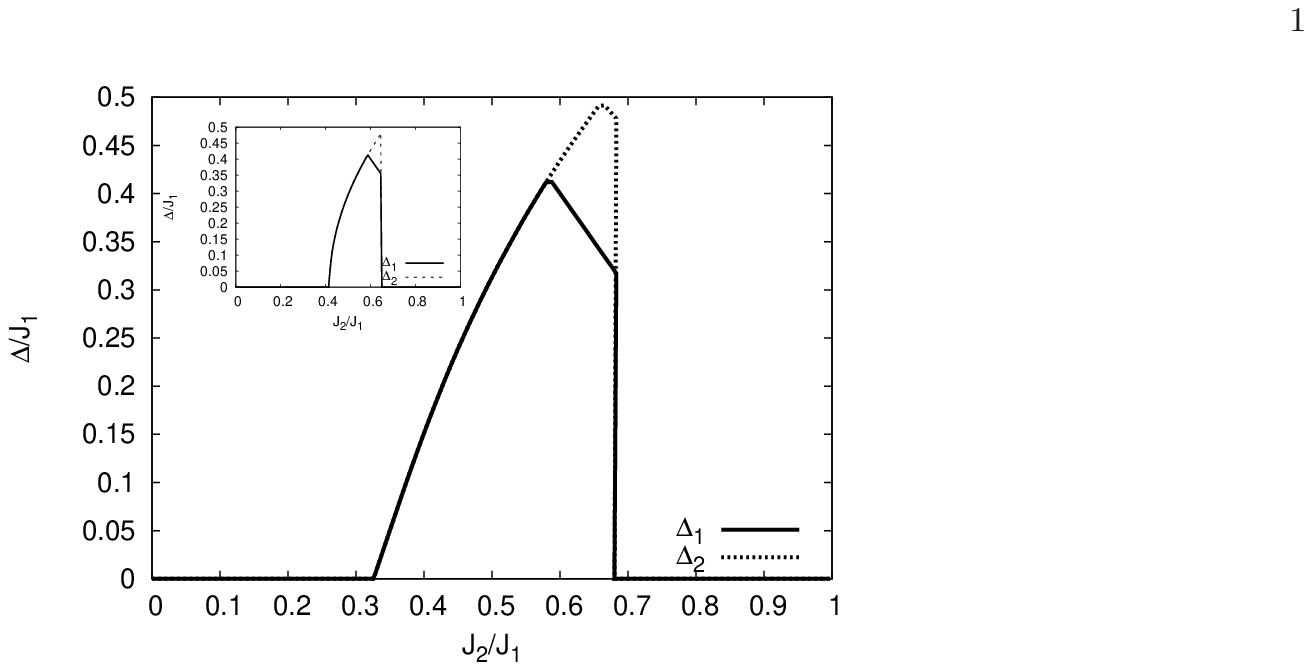}
 \end{center}
 \caption{The two lowest excitation energies taken at the center of the
	  plaquette Brillouin zone. The main panel shows the self-consistent
          solution to Bogoliubov's equations, while the inset corresponds to
          the time-dependent Gross-Pitaevskii equation (weak coupling). Since
          wavefunctions of collective excitations in the N\'eel and columnar
          phases have different symmetries, there are level crossings in the
          non-magnetic phase (cusps in the plot).}
 \label{fig_gaps}
\end{figure}
Expectation values of the functions Eq. (\ref{order_parameters}) in the
HF ground state are shown in Fig. \ref{fig_order_parameters}. Both phase
transition points $J_2^{c1}$ and $J_2^{c2}$ are clearly seen from this
plot. All functions change continuously across the second-order critical
point $J_2^{c1}$ and jump at the first-order transition point
$J_2^{c2}$. Except in the columnar phase the values of $F_4$ are
everywhere exactly twice larger than those of $F_1$, which is an
indication of the unbroken four-fold rotational symmetry of the lattice
in these regions. In the columnar phase, on the other hand, this
symmetry is broken and the above relation does not hold. While in the
N\'eel and columnar phases nonlocal terms in Eq.
(\ref{order_parameters}) are important, in the paramagnetic state the
only contribution to either expectation value comes from isolated
plaquettes (local terms in Eq. (\ref{order_parameters})), or
superplaquettes. This observation is consistent with properties of the
ground state in the non-magnetic phase, discussed earlier in this
section.

As mentioned in the Introduction, any choice of degree of freedom breaks
explicitly the lattice translational symmetry $T$, with the result that
links in the lattice become inequivalent. Indeed, the functions of Eq.
(\ref{order_parameters}), defined on the links, have non-zero values
even in the AF phase at $J_2=0$. However, this effect vanishes in the
thermodynamic limit (i.e., as the size of the degree of freedom is
increased). The finite-size scaling for $F_4(J_2/J_1)$ is presented in
the inset to Fig. \ref{fig_order_parameters}. The three extrapolated
values: $F_4(0)=-0.075$, $F_4(0.504)=-0.245$ and $F_4(0.997)=-0.004$,
suggest that the ``link-wise'' translational invariance is restored in
the thermodynamic limit in the N\'eel and columnar phases, but not in
the plaquette crystal phase. Moreover, the value of the jump
$g\equiv\frac{J_1d^2E_0}{NdJ_2^2}\bigr|_{J_2^{c1}-0}^{J_2^{c1}+0}$,
extrapolated to the thermodynamic limit (see inset to Fig.
\ref{fig_gse_d2_2x2_4x4}) remains finite: $g(J_2^{c1})=-1.006$. In other
words, these results imply that the critical point $J_2^{c1}$ corresponds
to the usual Landau second-order phase transition.

\subsection{Excitations in the plaquette crystal phase}

Until now we have considered only ground-state properties of the model
Eq. (\ref{hamiltonian}). However, low-lying excited states are also of
considerable interest.  In particular, the paramagnetic phase is known
to have gapped excitations, while N\'eel and columnar phases exhibit
Goldstone modes. Thus, the phase transition points $J_2^{c1}$ and
$J_2^{c1}$ must be accompanied by the opening of a gap in the excitation
spectrum: the former in a continuous and the latter in a discontinuous
fashion. In Appendix B we present a particular method to obtain the
collective spectrum of the system. The main idea of this approximation
is borrowed from the Bogoliubov-Fetter theory of superfluidity
\cite{Fetter_1972}. Namely, assume that on each plaquette the majority
of Schwinger bosons form a condensate in an appropriately chosen lowest
energy state and neglect fluctuations in the number of condensed
particles. We note, however, that due to the Schwinger boson constraint,
this quantity has the meaning of a probability to find a given plaquette
in the lowest energy HF state, rather than the number of particles.
Nevertheless, we will call it the condensate fraction $n_0$, which, in
principle, should be determined self-consistently, and is a measure of
the applicability of the entire approximation: it should satisfy the
inequality $|n_0-1|\ll1$. Once the condensation part is separated from
$\gamma_{ia}$, what remains describes fluctuation corrections to the HF
ground state. These fluctuations have rather strong effects near the
phase transition points, leading to the modification of $J_2^{c1}$ to a
$\lambda$-point and its considerable shift. The value of $J_2^{c2}$ also
changes, but much less significantly. These facts imply that our
approximation breaks down near the phase transition points. Indeed, in
Appendix B it is shown that close to the transition, the condensate is
strongly suppressed. However, deep in each phase $n_0\sim0.9$, thus
allowing us to draw conclusions about general properties of the
collective spectrum.

\begin{figure}[!t]
 \begin{center}
  \includegraphics[width=0.88\columnwidth]{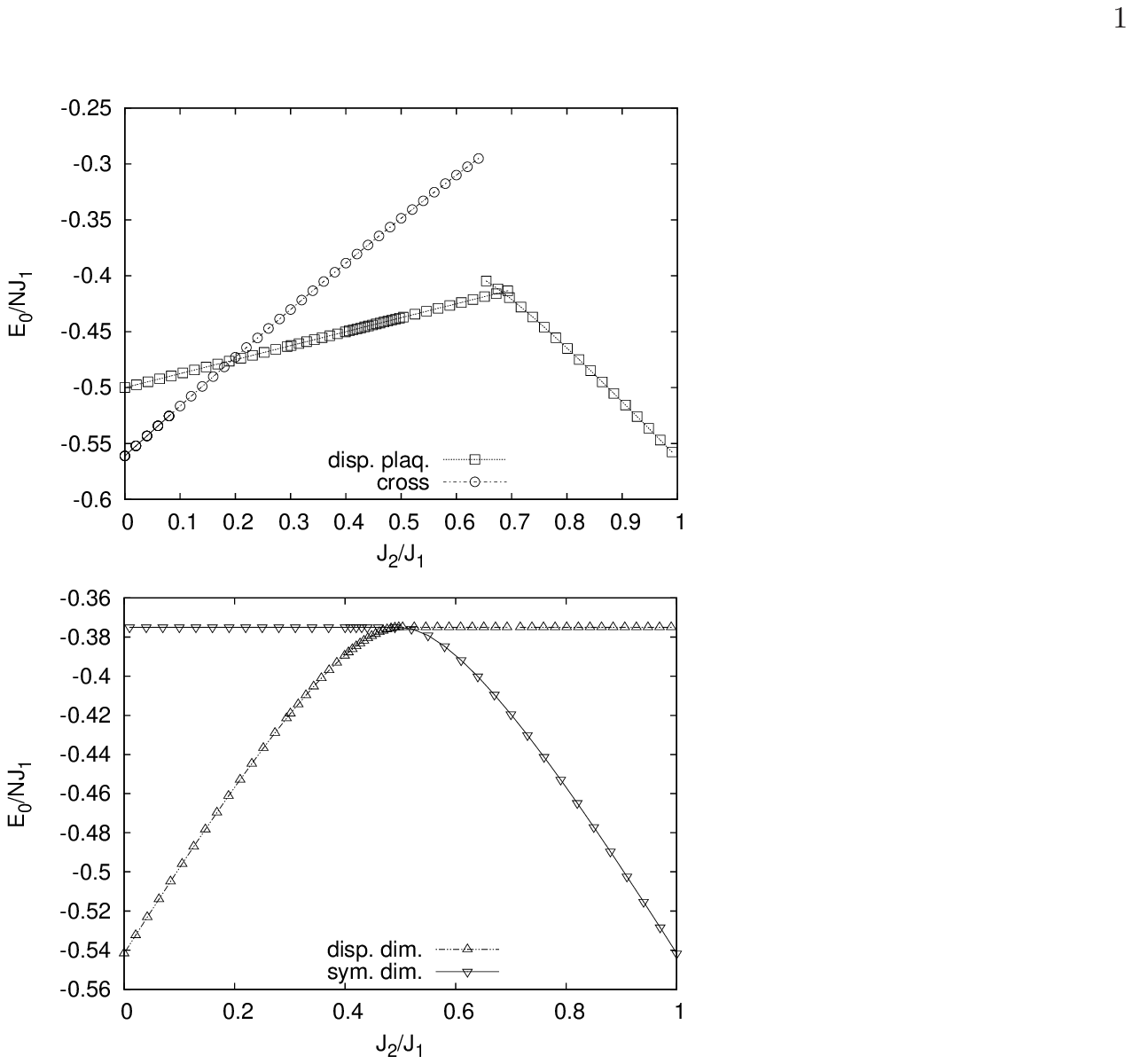}
 \end{center}
 \caption{HF ground state energy per spin for the displaced plaquette
	  and cross (upper panel), and dimer (lower panel) coverings.}
 \label{fig_gse_dp_dd_cr}
\end{figure}

The complete summary of the results is given in Appendix B, here we only
present the most interesting one: The gap in the excitation spectrum as
a function of $J_2/J_1$. Although we focus only on case (a) ($2\times2$
plaquette), the superplaquette degree of freedom can be considered in a
similar manner. The gap always occurs in momentum space at $\bk=0$,
which reflects translational invariance of the {\it plaquette} lattice.
Below we focus only on this point in the plaquette Brillouin zone. In
fact, there are $16-1=15$ collective branches and only some of them
become gapless in the phases with spontaneously broken $SU(2)$. However,
in the paramagnetic phase all branches develop a gap. In Fig.
\ref{fig_gaps} we show the energy gap $\Delta(J_2)=\omega(\bk=0,J_2)$
for the two lowest excitation branches in a system of $100\times100$
plaquettes, which approximates well the thermodynamic limit. The main
panel shows results of the self-consistent solution of Bogoliubov's
equations. The inset compares it with the solution of the time-dependent
Gross-Pitaeveskii equation (\ref{GP_equation}), which corresponds to the
weak-coupling approximation. In the N\'eel and columnar phases there are
two spin wave-type Goldstone modes, both of which acquire a gap in the
paramagnetic phase, at $J_2^{c1}$ and $J_2^{c2}$. However, as it follows
from Fig. \ref{fig_gaps}, positions of these points change from their HF
values to: $J_2^{c1}\approx0.33J_1$ and $J_2^{c2}\approx 0.65J_1$. The
critical point $J_2^{c1}$ was obtained by extrapolation of the staggered
magnetization curve (Fig. \ref{fig_mag_sf}) to zero, while the
first-order point $J_2^{c2}$ by extrapolating the two GSE curves in Fig.
\ref{fig_gse_sf} until intersection. The single-plaquette physical
picture, discussed previously in connection with the paramagnetic phase,
remains valid, e.g. the condensation occurs again in the plaquette state
$|1100\rangle$. Using this observation and symmetries of the matrix
elements $\bigr(H_{\rm int}^\sigma\bigr)^{a^\prime b^\prime}_{ab}$, one
can rigorously show the existence of a gap in the non-magnetic region.
In fact, we can say, that it is a property of our HMF approximation,
rather than a numerical evidence.

\subsection{Other degrees of freedom}

Finally, we comment on the results for cases (c)-(e), which, contrary to
the configurations considered before, explicitly break lattice
rotational symmetry, see Figs.
\ref{displ_plaquette_covering}-\ref{cross_covering}. The corresponding
GSEs are shown in Fig. \ref{fig_gse_dp_dd_cr}. In contrast to the
previously considered scenarios, these cases give {\it qualitatively}
wrong phase diagrams. Indeed, if we cover the lattice with  displaced
plaquettes or crosses, there exists no classical spin configuration,
which gives the long-range N\'eel order. On the other hand, such
configuration exists for the columnar state. For the displaced plaquette
covering the low-$J_2$ phase $J_2\le J_2^{c2}$ is an $SU(2)$ singlet,
and spatially is a set of non-interacting plaquettes (notice the
coincidence of $2\times2$--plaquette energies in the paramagnetic phases
of Figs. \ref{fig_gse_2x2_4x4} and \ref{fig_gse_dp_dd_cr}). Thus, the
phase transition to the columnar state is of the first-order. For the
cross covering, on the other hand, $SU(2)$ is explicitly broken for all
values of $J_2$, but since the columnar phase partially restores the
lattice translational invariance, it is again separated from the
non-magnetic state by a first-order phase transition point. The two
dimer configurations, case (d), are complementary to each other in the
sense that one of them has only the classical N\'eel state and another
-- only columnar phase. It follows that these configurations can have
only one second-order critical point at which $SU(2)$ is restored and
other symmetries remain broken. As a result, one obtains the phase
diagram, shown in the lower panel of Fig. \ref{fig_gse_dp_dd_cr}, which
is invariant under reflection in the plane $J_2=J_2^{c1}$. These
observations imply that the coarse graining prescriptions (c)-(e) are
probably a bad starting point for any approximation scheme.

\section{Discussion}

In this section we would like to put our main results in perspective by
making several summarizing remarks. It should be emphasized, that the
discussion below is based on our finite-size scaling results.

First of all, from our analysis it follows that dimer (bond) order is
always unfavorable in the non-magnetic phase. Notice that even when the
plaquette coverings were considered such an order did not occur,
although spontaneous dimerization was not  explicitly prohibited.
Instead, the quantum paramagnetic phase prefers to preserve the lattice
rotational symmetry, which makes the phase transition separating it from
the N\'eel phase fit perfectly well within the Ginzburg-Landau-Wilson
paradigm. The data presented for the staggered magnetization, Fig.
\ref{fig_mag_2x2_4x4}, and symmetry-breaking observables, Fig.
\ref{fig_order_parameters}, indicate that the symmetry group of the
N\'eel state is a subgroup of the symmetry group in the paramagnetic
phase, as both phases break $T$ and preserve $C_4$, but the latter also
preserves $SU(2)$. On the contrary, there is no such group-subgroup
relation between the paramagnetic and columnar antiferromagnetic phases.
Consequently, the transition between these two states is first order.
These observations are summarized in Fig. \ref{fig_phase_diagram}.
Indeed, starting from the known symmetry in the N\'eel state and
assuming validity of the Landau theory, one can unambiguously rule out
dimerized structures in the paramagnetic phase, since they break lattice
rotational symmetry. Therefore, our results do not favor the scenario
of deconfined quantum criticality, advocated in Refs.
\onlinecite{Sachdev_2004,Richter_2008}.
As already discussed in the previous section, our method 
explicitly breaks a particular lattice translational symmetry: The
ground state in Eq. (\ref{hf_wavefunction}) at $J_2=0$ (AF phase), is not
invariant under a $[11]$ lattice translation. One way to cure this
problem is to consider variational wavefunctions of
the ``resonating plaquette'' type:
\begin{displaymath}
 |\Psi\rangle=\bigl(1+T_{11}\bigr)|{\rm HF}\rangle,
\end{displaymath}
which restore that symmetry, 
with $T_{11}$ the translation operator along the $[11]$ direction in the
lattice. This state describes two resonating plaquette
configurations, shifted with respect to each other along [11]. 
Results of calculations using this wavefunction for systems up to
$6\times6$ spins indicate that for $J_2<J_2^{c1}$ the ground state has
long-range N\'eel order and is paramagnetic for $J_2\in(J_2^{c1},J_2^{c2})$.
The intermediate phase has a plaquette crystal order, but with partially
restored translational invariance. The phase transition at $J_2^{c1}$ is still
of the second-order, which is not surprising, as it can be described solely in
terms of the $SU(2)$ order parameter. However, based on these system
sizes, we can not definitively conclude whether this phase
transition remains of the second-order or becomes weakly first-order in the
thermodynamic limit.

Next, we observe that despite profound differences between the 2D and 1D
equivalent $J_1$-$J_2$ models, their non-magnetic phases present some
similarities. The one-dimensional model is known to be quasi-exactly
solvable \cite{MG} at the point $J_2=0.5J_1$ and exhibits a paramagnetic
ground state with short-range correlations for $J_2$ above the critical
value \cite{Okamoto_1992} $J_2^c\approx0.24J_1$ (however, due to the
peculiar physics in one dimension, the critical point $J_2^c$ is an
essential singularity and, therefore, not obviously accessible for the
HMF approximation of the type presented here). In this non-magnetic
region the ground state is doubly degenerate, corresponding to two
possible coverings of a 1D lattice with dimers, in accordance with the
Lieb-Schultz-Mattis theorem \cite{LSM}.  Unfortunately, a finite-size
scaling calculation for the gap between the lowest and first excited
energy levels, based on exact diagonalization of the $2\times2$ and
$4\times4$ clusters with periodic boundary conditions, does not provide a
definitive answer to the question on whether the ground state of the 2D
$J_1$-$J_2$ model becomes degenerate in the region $J_2^{c1}\le J_2\le
J_2^{c2}$. This is indeed what one would expect on the basis of a
generalization of the  Lieb-Schultz-Mattis theorem to higher space
dimensions (see, e.g. Ref. \onlinecite{Hastings_2004}). At the HF level,
it is true that different {\it plaquette} coverings of the lattice have
the same energy (simply because each plaquette is in its singlet ground
state). However, the total number of such configurations grows {\it
sub-exponentially} $\sim2^{\sqrt{N}}$, which should be contrasted with
the dimer covering problem, where this number is known
\cite{Fisher_1961} to be exponentially large. Based on this distinction,
one may speculate that if our plaquette picture is valid, there are not
enough different plaquette configurations for the paramagnetic phase to
become a spin liquid (i.e., a {\it resonating plaquette state}). This
statement, certainly, requires a separate investigation.

Finally, we emphasize that the main goal of this work is to investigate
the fundamental symmetries of the phases exhibited by the $J_1$-$J_2$
model. Although the energies presented for the $2\times2$ and $4\times4$
plaquette cases are different from those, obtained by more sophisticated
numerical methods, they can be systematically improved by considering
correlated trial wavefunctions or by using more complex methods, which
build upon the results reported here. However, we expect that
symmetry-wise our conclusions will remain unchanged.  One of such
methods suited for computing the phase diagram of the model
(\ref{hamiltonian}), which received significantly less attention,
amounts to applying the Wilson renormalization group procedure
\cite{Kruger_2006} and the density matrix renormalization group (DMRG)
method \cite{Leeuwen_2000}. Application of the latter faces serious
difficulties in 2D (see Ref. \onlinecite{White_1996} for a related
discussion) and requires a mapping of the 2D system onto a chain, which
introduces a certain bias to the final results \cite{Leeuwen_2000}.
While formulation of a practical and efficient DMRG approach to 2D
systems is yet to be developed, we note that our results may provide a
useful and rational initial input for such an algorithm.

\section{Conclusion}

In summary, we analyzed the phase diagram of the 2D $J_1$-$J_2$
model on a square lattice, focusing on symmetries of the various phases.

We showed that in this model the hierarchical language \cite{Ortiz_2003}
is defined by identifying the {\it plaquette as a relevant degree of
freedom}. Using an unbiased and manifestly symmetry-conserving
mean-field approach we compared it with several other possible
candidates: dimer and cross degrees of freedom, as well as different
ways to cover a 2D lattice with plaquettes. Our results indicate that
the plaquette (and superplaquette) covering, which preserves the lattice
rotational symmetry
has the best energy among considered coarse graining scenarios
and it is the only one to reproduce known facts, such as the
intermediate phase with gapped excitations, concerning the
phase diagram of the model, while other configurations fail to
exhibit all quantum phase transition points.

\begin{figure}[!t]
 \begin{center}
  \includegraphics[width=0.65\columnwidth]{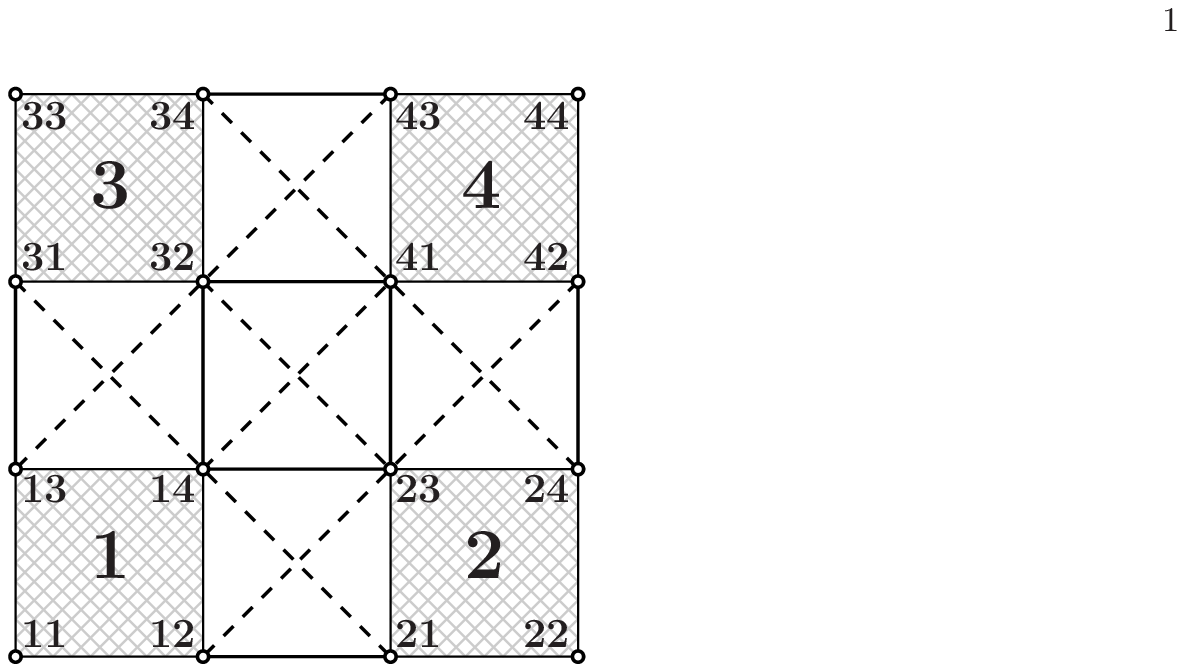}
 \end{center}
 \caption{The $4\times4$ superplaquette degree of freedom. Each spin
          carries two indices: a $2\times2$ plaquette number and a
          coordinate within this plaquette.}
 \label{fig_sp}
\end{figure}

Consistent with the previous work on the subject, we found the quantum
paramagnetic phase in the interval $0.42\le J_2/J_1\le0.66$. Main
controversies revolve around the nature of this intermediate
non-magnetic phase and of the quantum critical point separating it from
the N\'eel -- ordered state. We found that the paramagnetic phase is a
{\it plaquette crystal}, preserving both lattice and spin rotational
symmetries. Extrapolation of our numerical results to the thermodynamic limit
suggests that the Ginzburg-Landau-Wilson paradigm of phase transitions is
perfectly applicable in this case. Indeed, within the HMF, there is a
group-subgroup relation between symmetries of the non-magnetic
and N\'eel phases, which are thus separated by a second-order phase
transition. On the contrary, such relation does not exist between the
plaquette crystal and columnar antiferromagnetic phases, so the
corresponding transition is of the first order. Our plaquette crystal is
quite different from the usually proposed dimerized (bond-ordered)
phases in this non-magnetic region.

We also proposed a way to include fluctuations around the HF
ground state and showed that properties of the collective excitation
spectrum are consistent with the overall picture of spontaneously broken
symmetries. In particular, it was demonstrated that the quantum
paramagnetic state is characterized by a finite gap in the excitation
spectrum, which vanishes in the N\'eel and columnar phases, producing a
doubly degenerate Goldstone mode. 

Although currently there exists no known material whose ground state
realizes the paramagnetic phase of the $J_1$-$J_2$ model, in the future
momentum-resolved measurements, such as neutron diffraction methods, can
be used to identify the plaquette crystal phase of the type proposed
here. Its  experimental signature will be the unbroken four-fold lattice
rotational symmetry on both sides of the second-order phase transition
at the critical point $J_2^{c1}$.

We acknowledge fruitful discussions with C. D. Batista, C. Esebbag and
E. Fradkin. This work was  supported in part by the Spanish MEC under grant No.
FIS2006-12783-C03-01.

\section{Appendix A: Superplaquette degree of freedom}

Here we present details of the HF calculation which uses the $4\times4$
superplaquette, shown in Fig. \ref{fig_sp}, as an elementary degree of
freedom. It turns out that the full angular momentum basis is
inconvenient, thus we use the $2\times2$ plaquette product states in
order to perform the mean-field calculations. Each spin is characterized
by two indices: the plaquette number $i=1,\ldots,4$ and an index
$n=1\ldots4$, which specifies a vertex in the plaquette. The singlet
sector of the superplaquette Hilbert space is spanned by the states:
\begin{displaymath}
 |A\rangle=\sideset{}{^\prime}\prod_{i=1}^4|a_i\rangle,
\end{displaymath}
where prime indicates the constraint $\sum_{i=1}^4M_i=0$. Using these
states, we can write down matrix elements like
\begin{displaymath}
 \langle a^\prime_1\ldots a^\prime_4|\bS_{in}\bS_{jn^\prime}|a_1\ldots a_4
 \rangle
\end{displaymath}
in the compact form:
\begin{equation}
 \langle A^\prime|\bS_{in}\bS_{jn^\prime}|A\rangle=\bigl(\sigma_{nn^\prime}
 \bigr)_{a_ia_j}^{a_i^\prime a_j^\prime}\prod_{l\neq i,j}\delta_{a_l^\prime a_l}
 \label{average}
\end{equation}
with the symmetric matrices
$\bigl(\sigma_{nm}\bigr)_{a_ia_j}^{a_i^\prime a_j^\prime}=\bigl(\sigma_{mn}
\bigr)_{a_ja_i}^{a_j^\prime a_i^\prime}=\bigl(\sigma_{nm}\bigr)^{a_ia_j}_{
a_i^\prime a_j^\prime}$
defined as:
\begin{widetext}
 \begin{align}
  \bigl(\sigma_{nn^\prime}\bigr)_{a_ia_j}^{a_i^\prime a_j^\prime}=&
  \sum_{K,M}\bigl(-1\bigr)^{L_i+L_j^\prime+K}\langle L_i^\prime M_i^
  \prime L_j^\prime M_j^\prime|L_i^\prime L_j^\prime KM\rangle\langle L_iM_iL_j
  M_j|L_iL_jKM\rangle\times \nonumber \\
  &\qquad\qquad\times\left\{
  \begin{array}{ccc}
   L_i^\prime & L_j^\prime & K \\
   L_j & L_i & 1
  \end{array}
  \right\}\langle\lambda_i^\prime\|S_n\|\lambda_i\rangle
  \langle\lambda_j^\prime\|S_{n^\prime}\|\lambda_j\rangle, \nonumber
 \end{align}

 The Hamiltonian of a single superplaquette consists of two parts: a
 diagonal one, involving only $2\times2$ plaquette contributions and a
 non-diagonal part, which accounts for the plaquette interactions. The
 former is written down straightforwardly as a matrix:
 \begin{equation}
  \bigl(H_{\rm d}\bigr)_{A^\prime A}=\prod_i\delta_{a_i^\prime a_i}
  \sum_i\epsilon_{a_i},
  \label{se_sp_d}
 \end{equation}
 where $\epsilon_a$ is the plaquette self-energy, Eq.
 (\ref{plaquette_se}). The non-diagonal part has the operator form:
 \begin{align}
  H_{\rm nd}&=J_1\bigl[(\bS_{14}+\bS_{41})(\bS_{23}+\bS_{32})+\bS_{12}\bS_{21}+
  \bS_{13}\bS_{31}+\bS_{34}\bS_{43}+\bS_{24}\bS_{42}\bigr]+ \nonumber \\
  &+J_2\bigl[\bS_{12}\bS_{23}+\bS_{14}\bS_{21}+\bS_{23}\bS_{42}+\bS_{41}\bS_{24}
  +\bS_{34}\bS_{41}+\bS_{32}\bS_{43}+ \label{se_sp_nd} \\
  &\quad\qquad+\bS_{31}\bS_{14}+\bS_{13}\bS_{32}+\bS_{14}\bS_{41}+
  \bS_{32}\bS_{23}\bigr]. \nonumber
 \end{align}

 Let us now proceed with interaction terms in the $J_1$-$J_2$ Hamiltonian
 (\ref{hamiltonian}). Each superplaquette has 4 nearest and 4 next-nearest
 neighbors. Within each neighboring superplaquette we enumerate $2\times2$
 plaquettes by the indices 5, 6, 7, 8, so that $1\to5$, $2\to6$, $3\to7$ and
 $4\to8$. Enumeration of vertices within a plaquette stays the same. In this
 manner we have the symmetrized nearest neighbor
 \begin{align}
  H_1&=\frac{J_1}{4}\bigl[\bS_{11}(\bS_{73}+\bS_{62})+\bS_{12}\bS_{74}+
  \bS_{13}\bS_{64}+\bS_{31}\bS_{82}+\bS_{34}\bS_{52}+\bS_{33}(\bS_{84}+
  \bS_{51})+\nonumber \\
  &\quad\qquad+\bS_{21}\bS_{83}+\bS_{24}\bS_{53}+\bS_{22}(\bS_{84}+\bS_{51})+
  \bS_{43}\bS_{61}+\bS_{42}\bS_{71}+\bS_{44}(\bS_{73}+\bS_{62})\bigr]+
  \nonumber \\
  &+\frac{J_2}{4}\bigl[\bS_{11}(\bS_{74}+\bS_{64})+\bS_{12}(\bS_{73}+\bS_{83})+
  \bS_{13}(\bS_{62}+\bS_{82})+\bS_{21}(\bS_{74}+\bS_{84}) \label{sp_nn_int} \\
  &\quad\qquad+\bS_{22}(\bS_{53}+\bS_{83})+\bS_{24}(\bS_{51}+\bS_{71})+ 
  \bS_{31}(\bS_{84}+\bS_{64})+\bS_{33}(\bS_{82}+\bS_{52})+\nonumber \\
  &\quad\qquad+\bS_{34}(\bS_{51}+\bS_{61})+
  \bS_{42}(\bS_{73}+\bS_{53})+\bS_{43}(\bS_{62}+\bS_{52})+\bS_{44}(\bS_{71}+
  \bS_{61})\bigr], \nonumber
 \end{align}
\end{widetext}
and next-nearest neighbor
\begin{equation}
 H_2=\frac{J_2}{4}\bigl[\bS_{11}\bS_{84}+\bS_{22}\bS_{73}+\bS_{33}
 \bS_{62}+\bS_{44}\bS_{51}\bigr] \label{sp_nnn_int}
\end{equation}
superplaquette interactions. Using Eq. (\ref{average}), one can easily
construct matrix elements of the operators
(\ref{se_sp_nd})-(\ref{sp_nnn_int}), which are required to obtain the HF
equation of the type (\ref{HMF_equation}).

Having computed the single-superplaquette ground state wavefunction
$R^0_A=R^0_{a_1a_2a_3a_4}$, we can use it to determine the spin 
polarizations:
\begin{align}
 \langle {\rm HF}|S^z_{1n}|{\rm HF}\rangle=\bigl(S^z_n\bigr)_{a_1^\prime
 a_1}R&^0_{a_1^ \prime a_2a_3a_4}R^0_{a_1a_2a_3a_4}; \nonumber \\
 &\vdots \nonumber \\
 \langle {\rm HF}|S^z_{4n}|{\rm HF}\rangle=\bigl(S^z_n\bigr)_{a_4^\prime
 a_4}R&^0_{a_1a_2 a_3a_4^\prime}R^0_{a_1a_2a_3a_4}, \nonumber
\end{align}
where $\bigl(S^z_n\bigr)_{a^\prime a}$ is given by Eq. (\ref{spin_matr_el}).

\section{Appendix B: Fluctuation corrections -- superfluid mean-field}

In this Appendix we extend the analysis of Sec. III by considering
fluctuations around the HF ground state. While not unique, a natural way
to achieve this goal is to perform a superfluid-type mean-field
approximation. As a result one can obtain the collective spectrum and
corrections to the GSE and magnetization. Of primary interest is, of
course, the energy gap in the excitation spectrum.

The structure of the superfluid mean-field is similar to the
Fetter-Bogoliubov approach to inhomogeneous Bose liquids
\cite{Fetter_1972}. Although we shall present results only for
the $2\times2$
plaquette degree of freedom, it can equally be applied to the $4\times4$
superplaquette case.

\subsection{General formulation}

Let us return to the original Hamiltonian (\ref{ham_second_quant}) and 
explicitly separate out the condensate mode in the operators $\gamma_{ia}$:
\begin{equation}
 \gamma_{ia}=g_a+\beta_{ia}.
 \label{cwf}
\end{equation}
The condensation will occur in a certain superposition of the single-plaquette
states. The real-valued multiplet $g_a$ plays the role of a
condensate wavefunction (CWF) \cite{Fetter_1972}. Here it is chosen to
be spatially homogeneous, but inhomogeneous phases can also be included.
The CWF is normalized to the condensate fraction:
\begin{displaymath}
 \sum_ag_a^2=n_0.
\end{displaymath}
The non-condensate bosonic operators $\beta_{ia}$ describe fluctuation
corrections to the HF solution. If they are neglected, we naturally return to
the results of Sec. III. It is important to observe, however, that the HF
ground state corresponds to the Bose condensation on {\it each} lattice site,
not only in the $\bk=0$ mode.

The superfluid mean-field approximation amounts to enforcing the
Schwinger boson constraint on average:
\begin{equation}
 n_0+\frac{1}{N_\Box}\sum_{i,a}\langle\beta_{ia}^\dag\beta_{ia}\rangle=1,
 \label{part_cons}
\end{equation}
neglecting fluctuations in the condensate channel, and retaining only terms
quadratic in $\beta$ in the Hamiltonian (\ref{ham_second_quant}):
\begin{align}
 H&=N_\Box\biggl[\frac{1}{2}\biggl(\mu n_0+\sum_a\epsilon_ag_a^2\biggr)-
 \mu n_0\biggr]+\nonumber \\
 &+\sum_{i,a}(\epsilon_a-\mu)\beta_{ia}^\dag\beta_{ia}+4\sum_{i,\sigma}\bigl(
 H_{\rm int}^\sigma\bigr)^{a_1^\prime a_2^\prime}_{a_1a_2}g_{a_2^\prime}g_{a_2}
 \beta_{ia_1^\prime}^\dag\beta_{ia_1}+ \nonumber \\
 &+\sum_{\sigma,\langle ij\rangle_\sigma}\bigl(H_{\rm int}^\sigma\bigr)^{a_1^
 \prime a_2^\prime}_{a_1a_2}\bigl[g_{a_1^\prime}g_{a_2^\prime}(\beta_{ia_1}^
 \dag\beta_{ja_2}^\dag+\beta_{ia_1}\beta_{ja_2})+ \nonumber \\
 &\qquad\qquad\qquad\qquad+2g_{a_1}g_{a_2^\prime}\beta_{ia_1^\prime}^\dag
 \beta_{ja_2}\bigr],
 \label{Bog_hamiltonian}
\end{align}
where we abbreviated $\langle ij\rangle_\sigma=\bigl(\langle ij\rangle,
\langle\langle ij\rangle\rangle\bigr)$ and matrix elements of $H_{\rm
int}^\sigma$ are given by Eq. (\ref{interaction}). 

\begin{figure}[!t]
 \begin{center}
  \includegraphics[width=\columnwidth]{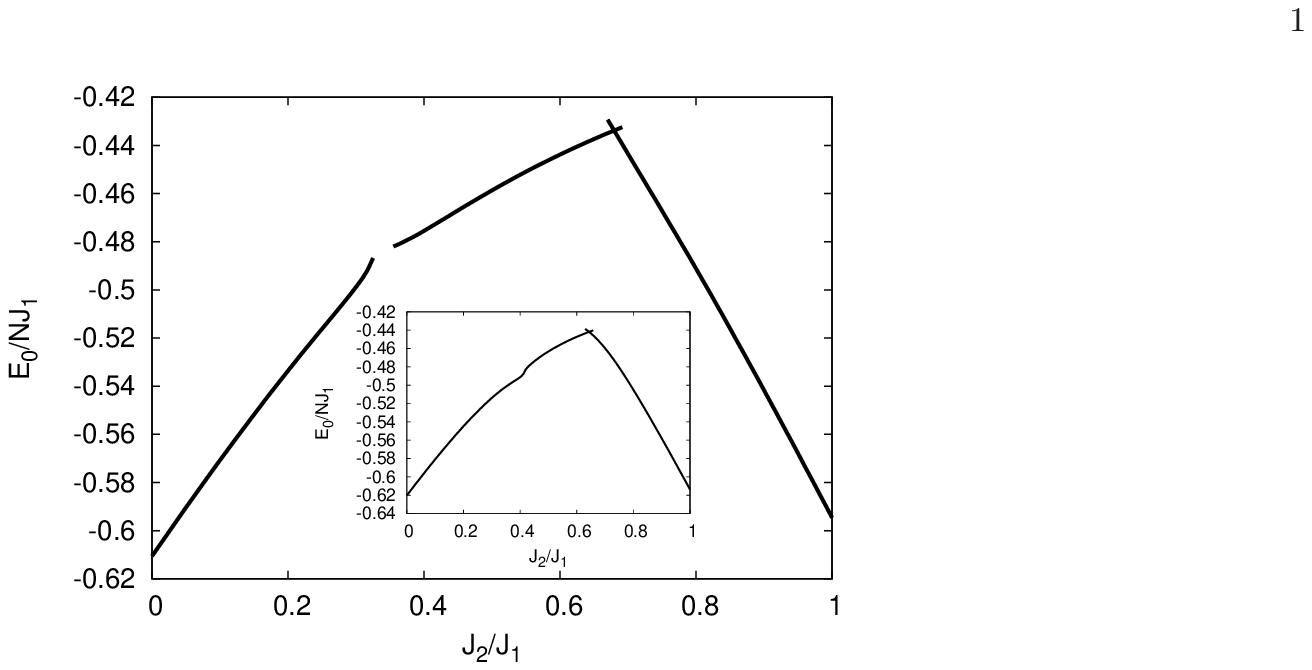}
 \end{center}
 \caption{Ground state energy for the self-consistent solution (main
	  panel) and after the first iteration (inset). The critical
	  point $J_2^{c1}$ becomes a $\lambda$-point. The absence of
	  points in the main panel around $J_2^{c1}$ is due to bad
	  convergence in the simulation.}
 \label{fig_gse_sf}
\end{figure}

The CWF $g_a$ is determined by the Gross-Pitaevskii equation, similar to the HF
equation (\ref{HMF_equation}):
\begin{equation}
 \biggl\{\epsilon_a\delta_{aa^\prime}+4\sum_\sigma\bigl(H_{\rm int}^\sigma
 \bigr)^{aa_1}_{a^\prime a_2}g_{a_1}g_{a_2}\biggr\}g_{a^\prime}=\mu g_a,
 \label{GP_equation}
\end{equation}
which defines the chemical potential $\mu$ and guarantees the
disappearance of linear terms in $\beta$ from the Hamiltonian of Eq.
(\ref{Bog_hamiltonian}). It is clear that $g_a(n_0=1)=R_a^0$ and
$\mu(n_0=1)=\varepsilon_0$. In other words, Eq. (\ref{GP_equation})
reproduces the results of Sec. III, if $n_0$ is forced to be unity.
Naturally, the first line in Eq. (\ref{Bog_hamiltonian}) coincides in
this limit (up to the chemical potential) with the expression
(\ref{HMF_energy}) for $E_0$. Quadratic terms in Eq.
(\ref{Bog_hamiltonian}) represent fluctuation corrections to the HF
results and constitute the focus of our analysis below.

\begin{figure}[!t]
 \begin{center}
  \includegraphics[width=\columnwidth]{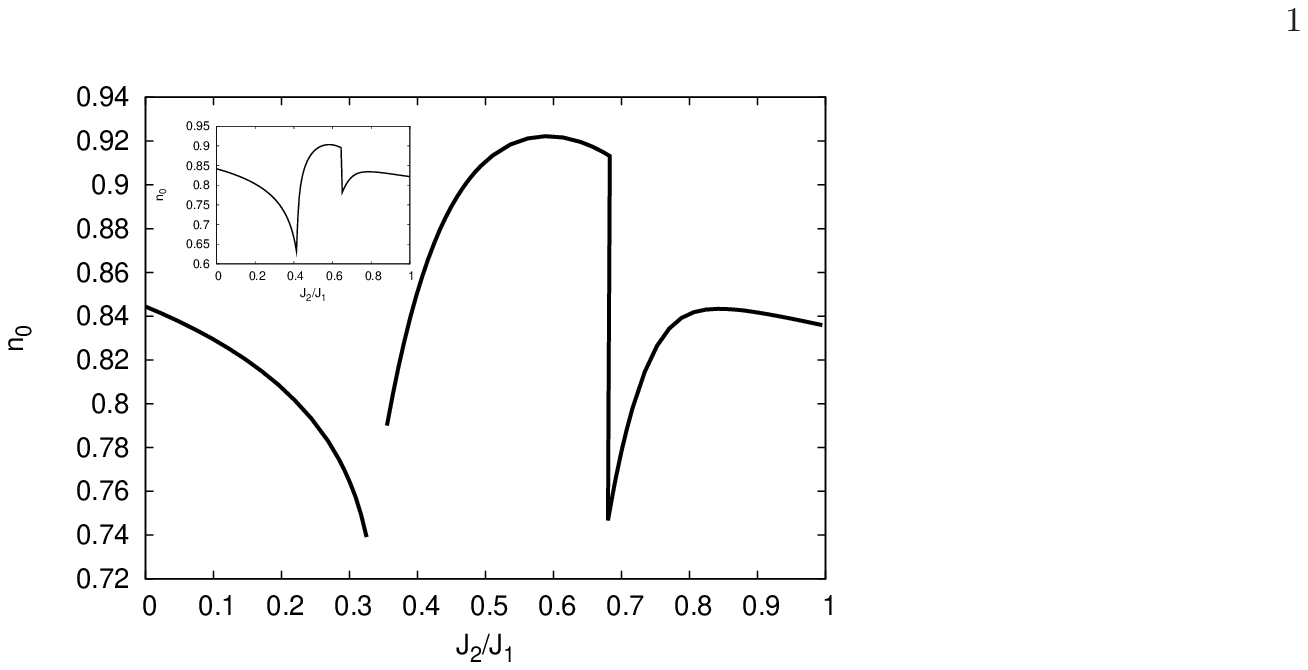}
 \end{center}
 \caption{Condensate fraction for the self-consistent solution (main
	  panel) and after the first iteration (inset). Notice the shift
	  of quantum phase transition points $J_2^{c1,2}$.}
 \label{fig_cf}
\end{figure}

The next step is to transform the quadratic part ($H_2$) of the Hamiltonian in
Eq. (\ref{Bog_hamiltonian}) into momentum space:
\begin{align}
 H_2&=\sum_{\bk,a}(\epsilon_a-\mu)\beta_{\bk a}^\dag\beta_{\bk a}
 +\sum_{\bk,\sigma}\bigl(H_{\rm int}^\sigma\bigr)^{a_1^\prime a_2^\prime}_{a_1
 a_2}\bigl\{\Theta_\bk^\sigma\bigl[g_{a_1^\prime}g_{a_2^\prime}\times
 \nonumber \\
 &\times(\beta_{\bk a_1}^\dag\beta_{-\bk a_2}^\dag+\beta_{\bk a_1}\beta_{\bk
 a_2})+2g_{a_1}g_{a_2^\prime}\beta_{\bk a_1^\prime}^\dag\beta_{\bk a_2}\bigr]+
 \nonumber \\
 &+4g_{a_2^\prime}g_{a_2}\beta_{\bk a_1^\prime}^\dag\beta_{\bk a_1}\bigr\},
 \label{Bog_hamiltonian_k}
\end{align}
where $\Theta_\bk^\sigma=(\cos k_x+\cos k_y,2\cos k_x\cos k_y)$ and $\bk$ is
defined within the plaquette Brillouin zone (i.e., there are $N_\Box$ $\bk$-
states). This Hamiltonian can be diagonalized by the Bogoliubov's
transformation:
\begin{align}
 \alpha_{\bk\nu}=&\sum_a\bigl(u_{\bk a}^\nu\beta_{\bk a}-
 v_{\bk a}^\nu\beta_{-\bk a}^\dag\bigr); \nonumber \\
 \alpha_{-\bk\nu}^\dag=&\sum_a\bigl(-v_{\bk a}^\nu\beta_{\bk a}+
 u_{\bk a}^\nu\beta_{-\bk a}^\dag\bigr),
 \label{Bog_transformation}
\end{align}
to a new set of bosonic operators $\alpha_{\bk\nu}$,
which represent quasiparticle excitations and annihilate the new ground
state: $\alpha_{\bk\nu}|\Psi_0\rangle=0$. Of course, only positive
quasiparticle energies, labeled by $\nu$, have physical meaning, however,
in order to obtain closure relations for the wavefunction
$\bigl(u_{\bk a}^\nu,v_{\bk a}^\nu\bigr)$ (which is, obviously, even in
$\bk$), we need to include zero-energy eigenvectors as well \cite{Ripka}.
\begin{figure}[!t]
 \begin{center}
  \includegraphics[scale=0.36,angle=-90]{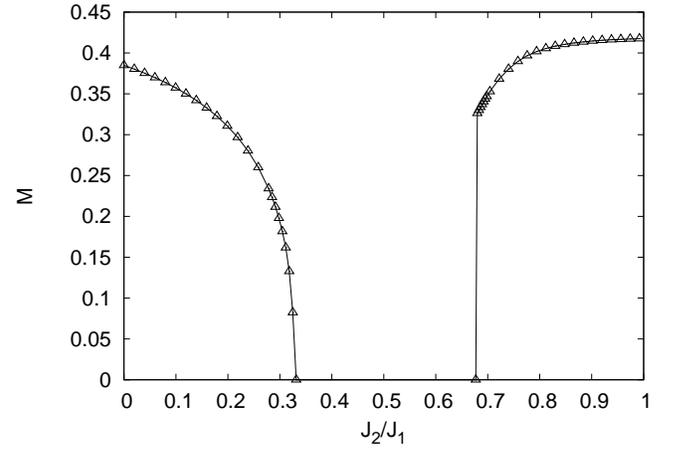}
 \end{center}
 \caption{Self-consistently computed staggered magnetization, $M_{\rm
	  stag}$, for $J_2\le J_2^{c1}$ and columnar magnetization along
	  the $x$-direction, $M_{\rm col}(x)$, for $J_2\ge J_2^{c2}$.}
 \label{fig_mag_sf}
\end{figure}
This completeness relation has the form, valid for {\it all} wavevectors:
\begin{align}
 \sum_\nu\bigl(u_{\bk a}^\nu u_{\bk b}^\nu-v_{\bk a}^\nu v_{\bk b}^\nu\bigr)=&
 \delta_{ab};  \nonumber \\
 \sum_\nu\bigl(u_{\bk a}^\nu v_{\bk b}^\nu-v_{\bk a}^\nu u_{\bk
b}^\nu\bigr)=&0.
 \label{completeness}
\end{align}
The amplitudes $u_a^\nu(\bk)$ and $v_a^\nu(\bk)$ are determined from
Bogoliubov's equations:
\begin{align}
 U_{ab}^N(\bk)u_{\bk b}^\nu+U_{ab}^A(\bk)v_{\bk b}^\nu=&\omega_\nu(\bk)
 u_{\bk a}^\nu; \nonumber \\
 U_{ab}^A(\bk)u_{\bk b}^\nu+U_{ab}^N(\bk)v_{\bk b}^\nu=&-\omega_\nu(\bk)
 v_{\bk a}^\nu,
 \label{Bog_eqs}
\end{align}
where we have introduced symmetric matrices:
\begin{align}
 U_{ab}^N(\bk)=&\frac{1}{2}\bigl(\epsilon_a-\mu\bigr)\delta_{ab}+\sum_\sigma
 \bigl(H_{\rm int}^\sigma\bigr)^{aa_1}_{a_2b}\Theta_\bk^\sigma g_{a_1}g_{a_2}+
 \nonumber \\
 &\quad+2\sum_\sigma\bigl(H_{\rm int}^\sigma\bigr)^{aa_1}_{ba_2}g_{a_1}g_{a_2};
 \label{U_matrices} \\
 U_{ab}^A(\bk)=&\sum_\sigma\bigl(H_{\rm int}^\sigma\bigr)^{a_1a_2}_{ab}
 \Theta_\bk^\sigma g_{a_1}g_{a_2}. \nonumber
\end{align}
It follows from Eq. (\ref{Bog_eqs}) that at each $\bk$ the quasiparticle
amplitudes obey the orthogonality conditions:
\begin{align}
 \sum_a\bigl(u_{\bk a}^\nu u_{\bk a}^{\nu^\prime}-v_{\bk a}^\nu 
 v_{\bk a}^{\nu^\prime}\bigr)=& \delta_{\nu\nu^\prime}; \nonumber \\
 \sum_a\bigl(u_{\bk a}^\nu v_{\bk a}^{\nu^\prime}-v_{\bk a}^\nu 
 u_{\bk a}^{\nu^\prime}\bigr)=&0.
 \label{orthogonality}
\end{align}
For any value of $\bk$ Bogoliubov's equations (\ref{Bog_eqs}) always
have at least two zero eigenvalues, which correspond to the zero-norm
eigenvector $u=-v=g$. This means that our case differs fundamentally
from the canonical superfluid Bose gas: instead of having a macroscopic
number of particle in one particular energy state, we obtain a
macroscopic number (equal to $N_\Box$) of condensation modes, each
containing less than one boson.

The quasiparticle energy equals $2\omega_\nu(\bk)$ and the GSE,
condensate fraction and spin polarization are expressed in terms of
$u_{\bk a}^\nu$ and $v_{\bk a}^\nu$ as:
\begin{align}
 \frac{E_0}{N}=&\frac{1}{8}\biggl(\mu n_0+\sum_a\epsilon_ag_a^2\biggr)+
 \frac{1}{4}\mu(1-n_0)- \nonumber \\
 &\quad-\frac{2}{N}\sideset{}{^\prime}\sum_{\bk,\nu,a}\omega_\nu(\bk)\bigl(
 v_{\bk a}^\nu\bigr)^2; \label{gse_cond_frac_mag} \\
 n_0=&1-\frac{1}{N_\Box}\sideset{}{^\prime}\sum_{\bk,\nu,a}\bigl(v_{\bk a}^\nu
 \bigr)^2; \nonumber \\
 \langle S^z_{in}\rangle=&\bigl(S^z_n\bigr)_{a^\prime a}\biggl[g_{a^\prime}g_a+
 \frac{1}{N_\Box}\sideset{}{^\prime}\sum_{\bk,\nu}v^\nu_{\bk a^\prime}
 v^\nu_{\bk a}\biggr]. \nonumber
\end{align}
In this expression $\bk$-summations are extended over the plaquette
Brillouin zone and $\nu$-summations over positive eigenvalues of Eq.
(\ref{Bog_eqs}), as indicated by the primes.

\begin{figure}[!t]
 \begin{center}
  \includegraphics[scale=0.34,angle=-90]{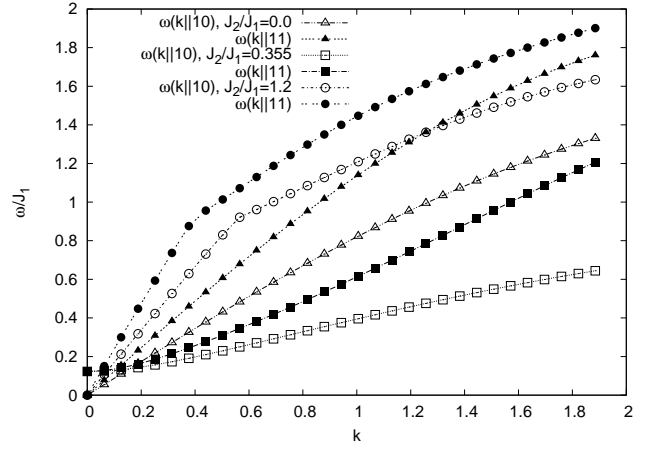}
 \end{center}
 \caption{The lowest excitation branch $\omega_1(\bk)$ along the $[10]$
	  and $[11]$ directions for three values of $J_2/J_1$ chosen in
	  different phases.}
 \label{fig_omega}
\end{figure}

\subsection{Results for the symmetric plaquette covering}

The condensate fraction $n_0$ should, in principle, be determined
self-consistently. The approximation is reasonable if $n_0\sim1$.
However, close to the phase transition points this is not true, since
fluctuations are very large in their vicinity. But deeply in each phase
the approximation works reasonably well, because $n_0$ turns out to be
of the order of $0.9$. Results of the numerical solution of Eqs.
(\ref{GP_equation}), (\ref{Bog_eqs}) and (\ref{gse_cond_frac_mag}) for
the symmetric covering of the lattice with $2\times2$ plaquettes are
shown in Figs. \ref{fig_gse_sf}-\ref{fig_omega}. The system size is
$100\times100$ plaquettes and periodic boundary conditions are assumed.
Figures' main panels correspond to the self-consistent solution and
their insets give results after the first iteration, which is equivalent
to solving the time-dependent Gross-Pitaevskii equation \cite{Ripka}.
Due to bad convergence close to the transition points (see, for
instance, Fig. \ref{fig_gse_sf}) the values of $J_2^{c1}$ and $J_2^{c2}$
were determined by extrapolation: $J_2^{c1}\approx0.33J_1$ and
$J_2^{c2}\approx0.65J_1$. The large shift of $J_2^{c1}$ compared to the
HF value is due to fluctuations in the $\beta$-channel, which renders
this point to be a $\lambda$-point, reduces the nominal value of the
magnetization in the N\'eel phase down to $M(J_2=0)\approx0.37$ (Fig.
\ref{fig_mag_sf}), and causes a great suppression of the condensate, as
shown in Fig. \ref{fig_cf}. 

However, the most interesting quantity to observe is the gap in the
excitation spectrum. Due to the homogeneity of the plaquette lattice, it
occurs at $\bk=0$ and is shown in Fig. \ref{fig_gaps}.

Technically, one may show that its very existence reflects the nature of
the ground state in the paramagnetic phase. Indeed, introducing linear
combinations of the amplitudes $u$ and $v$: $\varphi=u+v$ and
$\chi=u-v$, Bogoliubov's Eq. (\ref{Bog_eqs}) can be rewritten in the
form:
\begin{displaymath}
 \bigl(U^N+U^A\bigr)\bigl(U^N-U^A\bigr)\chi=\omega^2\chi.
\end{displaymath}
In the non-magnetic phase the condensation occurs in the lowest
plaquette state $|1100\rangle$: $g_a=\sqrt{n_0}\delta_{a,1100}$ and the
chemical potential coincides with its energy: $\mu=\epsilon_{1100}$.
Moreover, the matrix $\sum_\sigma\bigl(H_{\rm
int}^\sigma\bigr)_{b,1100}^{a,1100}$ vanishes. Writing down the
remaining matrices in (\ref{U_matrices}) at $\bk=0$, it is easy to see
that there exists only one vector $\chi$, which is annihilated by
$\bigl(U^N-U^A\bigr)$. Outside the intermediate region this simple
situation is not valid and there exist three eigenvectors $\chi$, which
correspond to $\omega^2=0$. One of them is the condensate mode and
should be discarded. The other two give doubly degenerate Goldstone
modes in the N\'eel and columnar phases. Here the self-consistent field,
determined by $g_a$, breakes the spin-rotational symmetry of the
original Hamiltonian. However, since the CWF $g_a$ belongs to the $M=0$
subspace, the generator $S_z$ remains an integral of motion. Thus, there
should be two Goldstone modes associated with rotations around the $x$
and $y$ axes \cite{Ripka}.

Our approximation correctly describes the excitation spectrum only at
small $\bk$. However, this is more than enough to observe that the
collective modes are of the spin-wave type in the N\'eel and columnar
phases, while in the paramagnetic phase the excitation band is
parabolic. These conclusions are summarized in Fig. \ref{fig_omega},
where we show the lowest branch $\omega_1(\bk)$ along two directions
$\bk\|[10]$ and $\bk\|[11]$ for three values of $J_2/J_1$, chosen in
different phases.

\end{document}